\documentclass[twocolumn,showpacs,preprintnumbers,amsmath,amssymb,superscriptaddress,showkeys,pra]{revtex4}

\usepackage{graphicx}
\usepackage{dcolumn}
\usepackage{bm}

\begin{document}

\preprint{APS/123-QED}

\title{Dynamic polarization effects on the angular distributions of protons \\
channeled through carbon nanotubes in dielectric media}

\author{D. Borka}
\email[Corresponding author:]{dusborka@vin.bg.ac.yu}
\affiliation{Laboratory of Physics (010), Vin\v ca Institute of Nuclear Sciences, P.O. Box 522\\
11001 Belgrade, Serbia}
\affiliation{Department of Applied Mathematics, University of Waterloo\\
Waterloo, Ontario, Canada N2L3G1}

\author{D. J. Mowbray}
\affiliation{Department of Applied Mathematics, University of Waterloo\\
Waterloo, Ontario, Canada N2L3G1}

\author{Z. L. Mi\v{s}kovi\' c}
\affiliation{Department of Applied Mathematics, University of Waterloo\\
Waterloo, Ontario, Canada N2L3G1}

\author{S. Petrovi\' c}
\affiliation{Laboratory of Physics (010), Vin\v ca Institute of Nuclear Sciences, P.O. Box 522\\
11001 Belgrade, Serbia}

\author{N. Ne\v{s}kovi\' c}
\affiliation{Laboratory of Physics (010), Vin\v ca Institute of Nuclear Sciences, P.O. Box 522\\
11001 Belgrade, Serbia}

\date{February 5, 2008}

\begin{abstract}
The best level of ordering and straightening of carbon nanotube
arrays is often achieved when they are grown in a dielectric matrix,
so such structures present the most suitable candidates for future
channeling experiments with carbon nanotubes. Consequently, we
investigate here how the dynamic polarization of carbon valence
electrons in the presence of various surrounding dielectric media
affects the angular distributions of protons channeled through
(11,~9) single-wall carbon nanotubes. Proton speeds between 3 and 10
a.u., corresponding to energies of 0.223 and 2.49 MeV, are chosen
with the nanotube's length varied between 0.1 and 1 $\mu$m. We
describe the repulsive interaction between a proton and the
nanotube's atoms in a continuum-potential approximation based on the
Doyle-Turner potential, whereas the attractive image force on a
proton is calculated using a two-dimensional hydrodynamic model for
the dynamic response of the nanotube valence electrons, while
assigning to the surrounding medium an appropriate (frequency
dependent) dielectric function. The angular distributions of
channeled protons are generated using a computer simulation method
which solves the proton equations of motion in the transverse plane
numerically. Our analysis shows that the presence of a dielectric
medium can strongly affect both the appearance and positions of
maxima in the angular distributions of channeled protons.
\end{abstract}



\maketitle

\section{Introduction}

While progress in theoretical modeling and computer simulations of
ion channeling through carbon nanotubes has reached a mature level,
as recently reviewed in Ref. \onlinecite{misk07}, efforts of
experimentalists have only recently begun to bear fruit in this
important research area. Because the best level of ordering and
straightening of carbon nanotubes is achieved when they are grown in
a dielectric matrix, such structures are perhaps the most suitable
candidates for ion channeling through carbon nanotubes. It thus came
as no surprise when Zhu \emph{et al.} \cite{zhu05} recently reported
the first experimental data on He$^{+}$ channeling through an array
of well ordered, multi-wall carbon nanotubes (MWNTs) which were
grown in a porous anodic aluminum oxide
(Al$_\mathrm{2}$O$_\mathrm{3}$) membrane. On the other hand, carbon
nanotubes have also been grown selectively within etched ion tracks
in SiO$_\mathrm{2}$ layers on Si by another experimental group
\cite{berd}, thus offering an interesting possibility for
realization of ion channeling through individual, single-wall carbon
nanotubes (SWNTs) at a wide range of ion energies. In addition, in
many applications of carbon nanotubes it is desirable to have them
embedded in a dielectric such as SiO$_\mathrm{2}$ \cite{tset06}, or
clamped by a metal shield \cite{misk07} made of nickel, which is
known to most readily bind to carbon nanotubes \cite{guer94}.

For ion channeling at the low (keV) and high (GeV) ends of the
energy range, the surrounding material would predominantly serve as
a passive container of carbon nanotubes when the dynamics of ion
motion is concerned. However, ions moving with medium (MeV) energies
will induce strong dynamic polarization of valence electrons in the
nanotubes which in turn will give rise to a sizeable image force on
the ions, as well as a considerable energy loss due to the
collective, or plasma, electron excitations. The dynamic image force
has been recently shown to give rise to the rainbow effect in the
angular distributions of protons channeled through short (11,~9)
single-wall \cite{borka06} and double-wall carbon nanotubes
\cite{borka07} in free space, which is not otherwise observable in
simulations of ion channeling through chiral carbon nanotubes using
the continuum approximation for the interaction potential
\cite{misk07}. Obviously, the presence of dielectric media may
affect these dynamic polarization forces, as well as the resulting
ion trajectories, making the analysis of such effects in ion
channeling through carbon nanotubes a timely task, which we take up
in this contribution. In particular, we are interested here
primarily in the effects of dielectric media on the dynamic image
force on channeled ions and their angular distributions.

The rainbow effect occurs and plays an important role in photon
scattering from water droplets \cite{khar74nuss79,jack99},
nucleus-nucleus collisions \cite{ford59,mcvo86,mich02}, atom or ion
collisions with atoms or molecules \cite{conn81}, electron-molecule
collisions \cite{zieg87}, atom or electron scattering from crystal
surfaces \cite{kley91,rein94}, and ion channeling in crystals
\cite{nesk86, krau86krau94}. Moreover, the rainbow effect has been
investigated recently in the context of grazing scattering of atoms
from metal surfaces under channeling conditions by Sch\"{u}ller
\emph{et al.} \cite{schu04schu05} who showed that precise
measurements of the well-defined maxima in the angular distributions
of scattered atoms, attributed to the rainbow effect, can give
detailed information on the interaction potential of the atoms with
the metal surfaces. On the other hand, the theory of crystal
rainbows has been formulated as the proper theory of ion channeling
in thin crystals \cite{petr00}, and has been subsequently applied to
ion channeling in short carbon nanotubes
\cite{petr05a,petr05b,borka05,nesk05}. It is therefore expected
that, in analogy with the surface channeling experiments
\cite{schu04schu05}, measurements of the rainbow effect in carbon
nanotubes may give precise information on both the atomic
configuration and the interaction potentials within such structures,
which are not completely known at present.

Previously reported simulations of ion channeling in carbon
nanotubes \cite{artr05,bell05,kras05} paid virtually no attention to
the effects of dynamic polarization of the nanotube valence
electrons. However, this process is expected to contribute to the
ion energy loss and to give rise to strong image forces on the
medium-energy ions \cite{mowb04mowb04}, as was recently demonstrated
in the computer simulations of angular distributions of protons
channeled through chiral single-walled carbon nanotubes in vacuum
\cite{zhou05}. The importance of the image force has also been
emphasized in the related area of ion transmission through
cylindrical channels in metals
\cite{aris01aris01,toke00,toke01,toke05,yama07}. Whereas interesting
parallels can be drawn between ion channeling through carbon
nanotubes and ion transmission through capillaries, it is important
to notice several crucial differences. Namely, while the lengths of
capillaries in metals can be comparable to those of nanotubes
considered here for ion channeling, their diameters are typically an
order of magnitude, or more, larger than those of carbon nanotubes,
making their aspect ratios considerably smaller. More importantly,
the inner surfaces of such broad channels in metals usually appear
to be quite rough, which does not seem to affect too much the
transmission of slow, highly-charged ions through such structures
\cite{toke00,toke01}. However, it is questionable whether conditions
required for channeling of fast ions leading to the rainbow effect
can be met in capillaries with such characteristics
\cite{schu04schu05,wint02}. On the other hand, in cases where carbon
nanotubes are grown in amorphous channels in a dielectric such as
Al$_\mathrm{2}$O$_\mathrm{3}$ \cite{zhu05} and SiO$_\mathrm{2}$
\cite{berd}, or are coated by an amorphous layer of metal
\cite{guer94,chai07}, it is precisely the regular atomic structure
of carbon nanotube that acts as a smooth "sleeve", or "mantle"
covering the underlying rough surface of the surrounding material,
thus enabling ion channeling through such structures. It remains to
be seen, however, whether large, multi-walled carbon nanotubes can
be grown inside the broad capillaries in metals enabling some sort
of ion channeling in their interior hollow regions.

We shall investigate here how dynamic polarization of carbon valence
electrons influences the angular distributions of protons channeled
in (11,~9) SWNTs with different dielectric media surrounding the
nanotubes. We consider proton speeds between 3 and 10 a.u., and
nanotube lengths between 0.1 and 1 $\mu$m. The image force acting on
an ion moving in a nanotube surrounded by a dielectric medium has
been recently calculated by means of a two dimensional (2D)
hydrodynamic model for the carbon valence electrons, while the
surrounding medium was described by a suitable frequency dependent
dielectric function \cite{mowb06,mowb07}. In the present
simulations, we shall use the van der Waals radius of a carbon atom
(0.17 nm) to approximate the distance between the nanotube wall
and various dielectric media \cite{hulm04,hulm03}. This value also
agrees with our density functional theory (DFT) calculations of the
average equilibrium separation between graphene and a Ni (111)
surface, based on the methodology described in Ref.
\onlinecite{sole02}. Other details of our simulation are similar to
those reported earlier \cite{borka06, borka07}.

After outlining the basic theory used in modeling the dynamic
polarization effects of carbon nanotubes, we shall discuss the
results of our ion trajectory simulations and give our concluding
remarks. Atomic units will be used throughout unless explicitly
stated otherwise.

\section{Theory}

The system under investigation is a proton moving through an (11,~9)
SWNT surrounded by a dielectric medium. The z-axis coincides with
the nanotube axis and the origin lies in its entrance plane. The
initial proton velocity vector, $\vec{v}$, is taken to be parallel
to the z-axis. Proton speeds between 3 and 10 a.u., corresponding to
energies of 0.223 and 2.49 MeV, are considered, with the nanotube's
length varied between 0.1 and 1 $\mu$m.

We also assume that the repulsive interaction between the proton and
the nanotube atoms may be treated classically, using the
Doyle-Turner expression for the proton-carbon atom interaction
potential \cite{doyl68}, averaged axially \cite{lind65} and
azimuthally \cite{zhev98zhev00}. The repulsive potential for proton
channeling through the nanotube is then of the form

\begin{eqnarray}
U_\mathrm{rep}(r) & = & {\frac{{16\pi dZ_\mathrm{1} Z_\mathrm{2} }}{{3 \sqrt{3} l^{2}}} }\nonumber\\
& \times & {\sum\limits_\mathrm{j = 1}^{4} {a_\mathrm{j}
b_\mathrm{j}^{2} I_\mathrm{0} (b_\mathrm{j}^{2} rd) \exp \{ -
b_\mathrm{j}^{2}}} [r^{2} + (d / 2)^{2}]\},
\label{equ01}
\end{eqnarray}

\noindent where $Z_\mathrm{1}$ = 1 and $Z_\mathrm{2}$ = 6 are the
atomic numbers of proton and carbon atoms, respectively, $d$ = 2$a$
is the nanotube diameter, $l$ is the C-C bond length, $r$ is the
distance between the proton and nanotube axis, $I_\mathrm{0}$ is the
modified Bessel function, and $a_\mathrm{j} = \{ 0.115, 0.188,
0.072, 0.020 \}$ and $b_\mathrm{j} = \{ 0.547, 0.989, 1.982, 5.656
\}$ are fitting parameters in atomic units \cite{doyl68}.

The dynamic polarization of the nanotube is treated by a 2D
hydrodynamic model of the nanotube valence electrons, based on a
jellium-like description of the ion cores on the nanotube wall
\cite{mowb04mowb04,mowb06}. This model includes both axial and
azimuthal averaging consistent with our treatment of the repulsive
interaction. The self energy, or the image potential,
$E_\mathrm{self}$, for a single ion of charge $Z_\mathrm{1}$ at
position $\vec{r}_\mathrm{0}(t)$ is defined by

\begin{equation}
E_\mathrm{self} = ({{Z_\mathrm{1}}  \mathord{\left/ {\vphantom
{{Z_\mathrm{1}} {2}}} \right. \kern-\nulldelimiterspace}
{2}})\Phi_\mathrm{ind} (\vec{r_\mathrm{0}} (t), t),
\label{equ02}
\end{equation}

\noindent where $\Phi_\mathrm{ind} (\vec{r},t)$ is the potential at
the point $\vec r$, given in cylindrical coordinates by $\vec r = \{
{r,\varphi ,z} \}$, which is induced in the system by the presence
of ion. After performing the Fourier transform with respect to time,
and following the method of Doerr and Yu \cite{doer04}, we consider
the total electric potential to be the sum of the external
perturbing potential, $\Phi_\mathrm{ext} (\vec{r},\omega)$, and the
induced potential, $\Phi_\mathrm{ind} (\vec{r},\omega)$, due to
polarization of the nanotube and the dielectric boundary by the
proton,  so that

\begin{equation}
\Phi (\vec {r},\omega ) = \Phi _\mathrm{ext} (\vec {r},\omega ) +
\Phi_\mathrm{ind} (\vec {r},\omega ).
\label{equ03}
\end{equation}

\noindent The Poisson equation then gives

\begin{equation}
\nabla ^{2}\Phi _\mathrm{ext} (\vec {r},\omega ) = - 4\pi
\rho_\mathrm{ext} (\vec{r},\omega ),
\label{equ04}
\end{equation}

\begin{equation}
\nabla ^{2}\Phi _\mathrm{ind} (\vec {r},\omega ) = 4\pi
[n_\mathrm{1} (\vec {r}_\mathrm{a} ,\omega )\delta (r - a) -
\sigma_\mathrm{b} (\vec {r}_\mathrm{b} ,\omega )\delta (r - b)],
\label{equ05}
\end{equation}

\noindent where $\vec{r}_\mathrm{a}$ is a position on the nanotube
of radius $a$ given by $\vec{r}_\mathrm{a}$ = $\{ a, \varphi, z \}$,
$\vec{r}_\mathrm{b}$ is a position on the boundary of the dielectric
of radius $b$ given by $\vec{r}_\mathrm{b}$ = $\{ b, \varphi, z \}$,
$n_\mathrm{1} (\vec {r}_\mathrm{a}, \omega)$ is the induced electron
number density (per unit area) on the nanotube, and
$\sigma_\mathrm{b}(\vec{r}_\mathrm{b},\omega)$ is the polarization
charge density (per unit area) induced on the boundary of the
dielectric.

We may denote the Fourier transform in cylindrical coordinates of an
arbitrary function $A(r, \varphi, z, \omega )$ by

\begin{equation}
A(r,\varphi ,z,\omega ) = {\sum\limits_\mathrm{m} {\int
{{\frac{{dk}}{{\left( {2\pi} \right)^{2}}}}e^{im\varphi}
e^{ikz}\tilde {A}(r,m,k,\omega )}}}.
\label{equ06}
\end{equation}

\noindent The Green's function in cylindrical coordinates is then

\begin{equation}
{\frac{{1}}{{{\left\| {\vec {r} - \vec {{r}'}} \right\|}}}} =
{\sum\limits_\mathrm{m} {\int {{\frac{{dk}}{{\left( {2\pi}
\right)^{2}}}}e^{im(\varphi - {\varphi} ')}e^{ik(z -
{z}')}g(r,{r}',m,k)}} },
\label{equ07}
\end{equation}

\noindent where $g(r, r', m, k)$ is the radial Green's function,

$$
g(r,r',m,k) = 4 \pi I_\mathrm{m} (kr_\mathrm{<}) K_\mathrm{m}
(kr_\mathrm{>}),
$$

\noindent with $r_\mathrm{<} = \min \{ r,r' \}$, $r_\mathrm{>} =
\max \{ r,r' \}$, and $I_\mathrm{m}$ and $K_\mathrm{m}$ being the
modified Bessel's functions of the first and second kind,
respectively.

The Fourier transform of the external perturbing potential due to a
single proton of charge $Z_\mathrm{1}$ = 1 moving parallel to the
nanotube axis with constant speed $v$, such that
$\vec{r_\mathrm{0}}(t)$ = $\{ r_\mathrm{0}, \varphi_\mathrm{0}, vt
\}$, is given by

\begin{equation}
\tilde {\Phi}_\mathrm{ext}(r) = {\frac{{2\pi Z_\mathrm{1}}}
{{\varepsilon_\mathrm{nt} }}}g(r,r_\mathrm{0},m,k)\delta (\omega -
kv)e^{ - im\varphi_\mathrm{0}},
\label{equ08}
\end{equation}

\noindent where $\varepsilon_\mathrm{nt}$ is the background optical
dielectric constant for the nanotube (for which we use
$\varepsilon_\mathrm{nt}$ = 1), $m$, $k$ and $\omega$ are the
angular oscillation mode, longitudinal wave number and angular
frequency of an elementary excitation of the nanotube atoms valence
electrons treated as an electron gas. Note that one may set
$\varphi_\mathrm{0}$ = 0 because of the axial symmetry of our model
for nanotube. The Fourier transform of the induced potential is
given by \cite{mowb06}

\begin{eqnarray}
\tilde {\Phi} _\mathrm{ind}(r) = {\frac{{ - a\tilde {n}_\mathrm{1}}}
{{\varepsilon _\mathrm{nt} }}}[g(r,a,m,k) & + & b\Re g(r,b,m,k){g}'(b,a,m,k)] \nonumber\\
& + & b\Re g(r,b,m,k){\frac{{\partial \tilde {\Phi} _\mathrm{ext}}}
{{\partial r}}}{\left| {_\mathrm{b}} \right.},
\label{equ09}
\end{eqnarray}

\noindent where $g'(r,r',m,k) \equiv \frac{\partial }{{\partial
r}}g(r,r',m,k)$, and the Fourier transform of the induced electron
number density on the nanotube is

\begin{equation}
\tilde {n}_\mathrm{1} = {\frac{{\tilde {\Phi} _\mathrm{ext} (a) +
bg(a,b,m,k)\Re {\frac{{\partial \tilde {\Phi} _\mathrm{ext}}}
{{\partial r}}}{\left| {_\mathrm{b}} \right.}}}{{\chi ^{ - 1} +
{\frac{{a}}{{\varepsilon _\mathrm{nt}}} }[g(a,a,m,k) + b\Re
g(b,a,m,k){g}'(b,a,m,k)]}}}.
\label{equ10}
\end{equation}

\noindent Here we have defined the response function  of the
polarization charge due to the external electric field in the radial
direction on the outer boundary of the medium, with dielectric
function $\varepsilon_\mathrm{\omega}$, by

\begin{equation}
\Re \equiv {\frac{{\varepsilon _\mathrm{\omega}  -
\varepsilon_\mathrm{nt}}} {{4\pi [\varepsilon _\mathrm{nt} +
(\varepsilon_\mathrm{nt} - \varepsilon _\mathrm{\omega}
)kbI_\mathrm{m} (kb){K}'_\mathrm{m} (kb)]}}}.
\label{equ11}
\end{equation}

\noindent The response function, $\chi$, for the induced electron
density on the nanotube due to the total electric potential, defined
by $\tilde n_\mathrm{1}  = \chi \tilde \Phi$, is given in the 2D
hydrodynamic model by

\begin{equation}
\chi \equiv {\frac{{n_\mathrm{0} (k^{2} + {{m^{2}} \mathord{\left/
{\vphantom {{m^{2}} {a^{2})}}} \right. \kern-\nulldelimiterspace}
{a^{2})}}}}{{\alpha (k^{2} + {{m^{2}} \mathord{\left/ {\vphantom
{{m^{2}} {a^{2})}}} \right. \kern-\nulldelimiterspace} {a^{2})}} +
\beta (k^{2} + {{m^{2}} \mathord{\left/ {\vphantom {{m^{2}}
{a^{2})}}} \right. \kern-\nulldelimiterspace} {a^{2})}}^{2} - \omega
(\omega + i\gamma )}}},
\label{equ12}
\end{equation}

\noindent where $n_\mathrm{0}$ is the equilibrium number density of
all four carbon valence electrons ($n_\mathrm{0} \approx$ 0.428
a.u.), $\alpha$ = $\pi n_\mathrm{0}$, $\beta$ = 1/4, and we take the
limit $\gamma \to 0^{+}$ \cite{mowb06}.

We note that the above theory yields the self energy (image
potential), $E_\mathrm{self}(r_\mathrm{0})$, as a stationary
function in the proton's moving frame of reference, which only
depends on proton's radial position $r_\mathrm{0}$ within an
infinitely long nanotube. In that respect, it should be possible to
limit our considerations to nanotubes which are long enough so that
we may ignore the edge effects on the image potential at the
entrance and exit planes and, at the same time, short enough that
the total energy losses of channeled protons may be neglected.
Therefore, when calculating the dynamic image force on a proton one
may safely consider its longitudinal velocity component as constant
equal to its initial speed $v$, while changes in the perpendicular
components of the proton velocity $\vec u$ can be neglected under
channeling conditions. Consequently, one may consider the radial
position $\vec{r}$ of a channeled proton to evolve adiabatically
under the action of an axially symmetric force field, $F(r) =
F_\mathrm{rep}(r) + F_\mathrm{image}(r)$, where $F_\mathrm{rep} (r)
= - dU_\mathrm{rep} (r)/dr$ and $F_\mathrm{image} (r) = -
dE_\mathrm{self} (r)/dr$. Because our numerical calculations of
proton trajectories will be executed under the assumption of a
homogeneous, mono-directional beam of protons incident in the
direction parallel to the nanotube axis, the resulting simulation
code will be essentially two-dimensional (2D). The effects on proton
channeling coming from a divergent beam and non-parallel incidence
will be studied by a full 3D code in future reports.

It should be also mentioned that the assumption that proton charge
remains fixed at $Z_\mathrm{1}$ = 1, which is used in our channeling
simulations, needs careful consideration. Making analogy with the
analysis of experiments on grazing scattering of protons on an Al
surface \cite{wint91} and referring to the data on proton
transmission through carbon foils \cite{kreu82}, one may expect that
the $Z_\mathrm{1}$ = 1 charge state would be the most dominant
fraction for protons channeled in carbon nanotubes at the speeds in
excess of $v$ = 3. Nevertheless, simultaneous measurements of the
angular distributions of transmitted particles in different charge
states could reveal a wealth of information on both the image
interactions and the charge transfer processes taking place in short
nanotubes, in close analogy with the experiments done by Winter
\cite{wint02,wint91}. On the other hand, for ion channeling through
longer nanoubes, the image force itself can be quite strongly
affected by the dynamics of the charge-changing events near the
nanotube wall, as was shown recently for grazing scattering of
protons from an Al surface \cite{misk05}. The effects of charge
transfer will be included in our future ion channeling simulations.

The angular distributions of transmitted protons are generated using
a Monte-Carlo computer simulation method. The Cartesian components
of the proton impact parameter, $x_\mathrm{0}$ and $y_\mathrm{0}$,
are chosen from a random uniform distribution within the nanotube
cross-sectional area. With the bond length between the nanotube
atoms being 0.144 nm \cite{sait01}, we obtain for the radius of the
(11,~9) nanotube of $a$ = 0.689 nm. Any protons with an impact point
inside the annulus with radii in the interval $[a -
a_\mathrm{sc},a]$, where $a_\mathrm{sc} = [9 \pi^{2} / (128 \
Z_\mathrm{2})]^\frac{1}{3}a_\mathrm{0}$ is the screening length
(with $a_\mathrm{0}$ being Bohr's radius), are treated as if they
were backscattered and are disregarded from the simulation. The
initial number of protons used is 3~141~929.

The Cartesian components of the proton scattering angle, i.e., of
the deflection function, $\Theta_\mathrm{x}$ and
$\Theta_\mathrm{y}$, are obtained via the expressions
$\Theta_\mathrm{x}$ = $u_\mathrm{x}/v$ and $\Theta_\mathrm{y}$ =
$u_\mathrm{y}/v$, where $u_\mathrm{x}$ and $u_\mathrm{y}$ are the
Cartesian components of the final perpendicular velocity vector,
$\vec{u}$, obtained in our simulation of transmitted protons. It has
been demonstrated that the proton channeling in nanotubes can be
analyzed successfully via the corresponding mapping of the impact
parameter plane, the $x_\mathrm{0}y_\mathrm{0}$ plane, to the
scattering angle plane, the $\Theta_\mathrm{x}\Theta_\mathrm{y}$
plane. However, as the total interaction potential in the case under
consideration is axially symmetric and the incident protons are
moving parallel to the nanotube axis, the analysis of the mapping
may be reduced to the analysis of the scattering angle, $\Theta =
(\Theta_\mathrm{x}^{2} + \Theta_\mathrm{y}^{2})^\frac{1}{2}$, as a
function of the impact parameter, $r_\mathrm{0} = (x_\mathrm{0}^{2}
+ y_\mathrm{0}^{2})^\frac{1}{2}$. Therefore, we may take
$y_\mathrm{0}$ = 0, and analyze the deflection function
$\Theta_\mathrm{x}(x_\mathrm{0})$ only. The extrema of this function
are the rainbow extrema, and the corresponding singularities
appearing in the angular distribution of channeled protons are the
rainbow singularities \cite{borka06}.

\section{Results and discussion}

\begin{figure}
\centering
\includegraphics[width=0.47\textwidth]{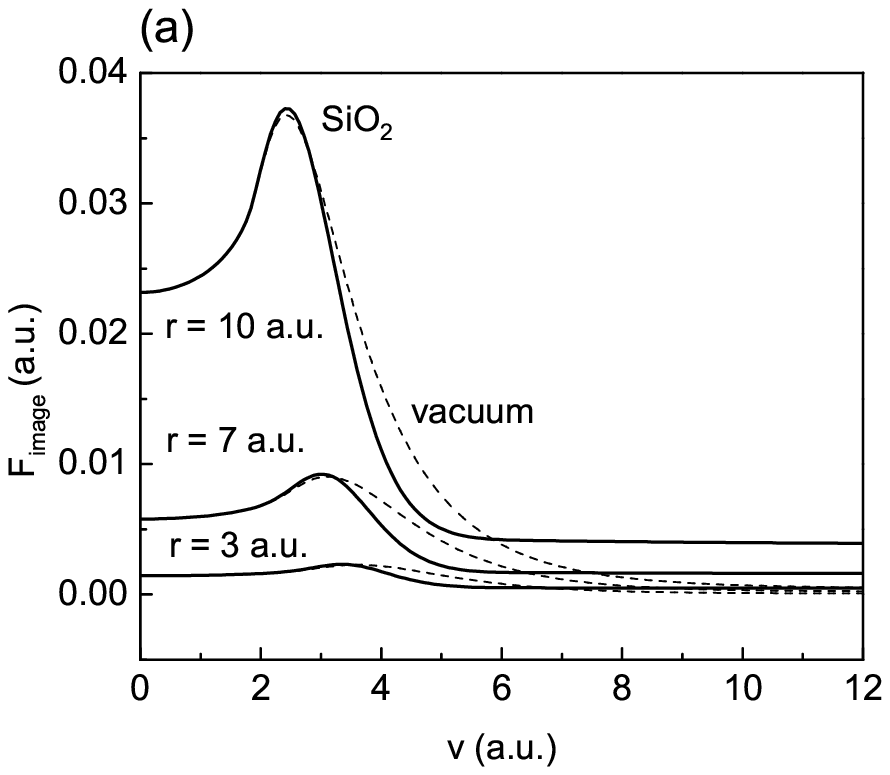}
\includegraphics[width=0.47\textwidth]{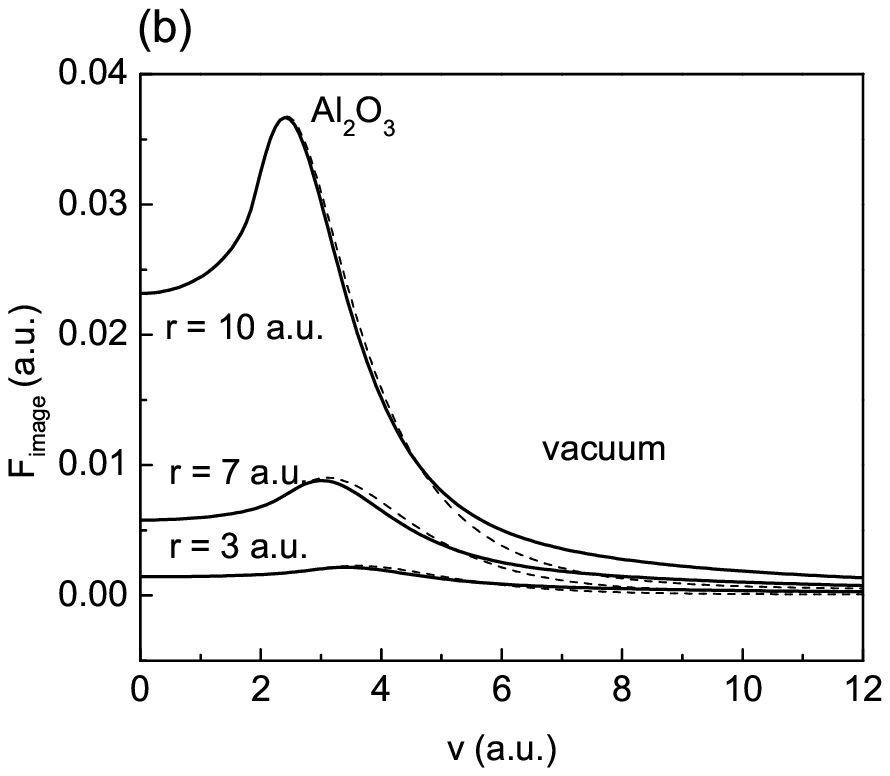}
\includegraphics[width=0.47\textwidth]{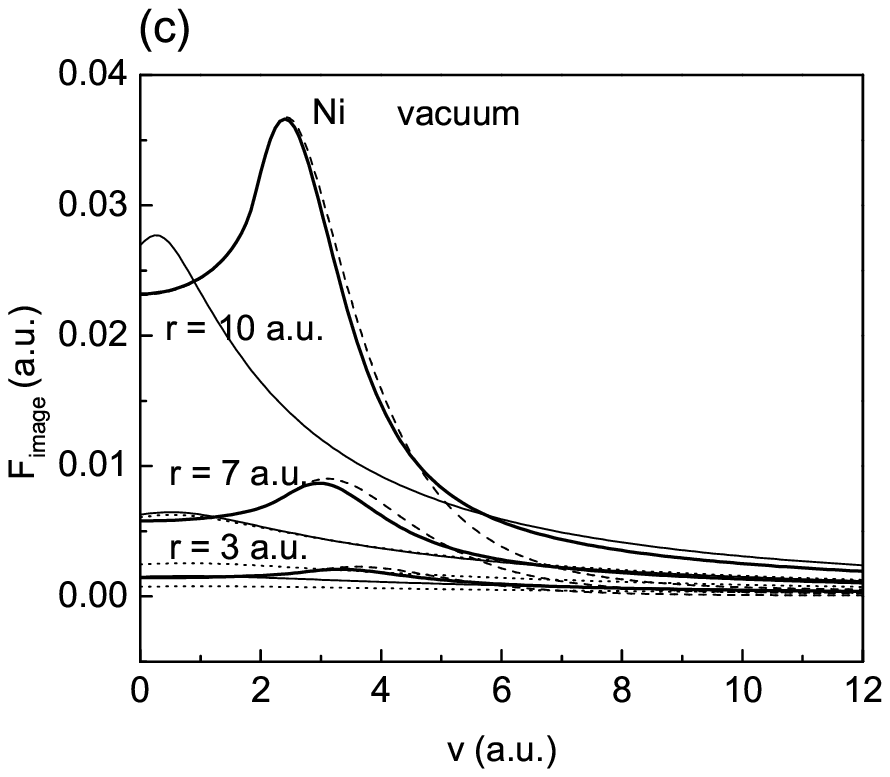}
\caption{The image force of a proton traveling paraxially at three
distances from the nanotube axis: $r$ = 3 (lower), 7 (middle) and 10
a.u. (upper), as a function of proton speed $v$ in a.u., due to an
(11,~9) SWNT in vacuum (dashed curves) and encapsulated by an (a)
SiO$_\mathrm{2}$ channel, (b) Al$_\mathrm{2}$O$_\mathrm{3}$ channel
and (c) Ni channel (solid curves). The radius of the nanotube is $a$
= 13.01 a.u. and the radius of the dielectric channel is $b$ = 16.22
a.u. In panel (c) we also show the image force for the same three
distances inside Ni channels without encapsulated nanotube for two
channel radii: $b$ = 13.01 a.u. (thin solid lines) and $b$ = 16.22
a.u. (dotted lines).} \label{fig01}
\end{figure}

\begin{figure}
\centering
\includegraphics[width=0.46\textwidth]{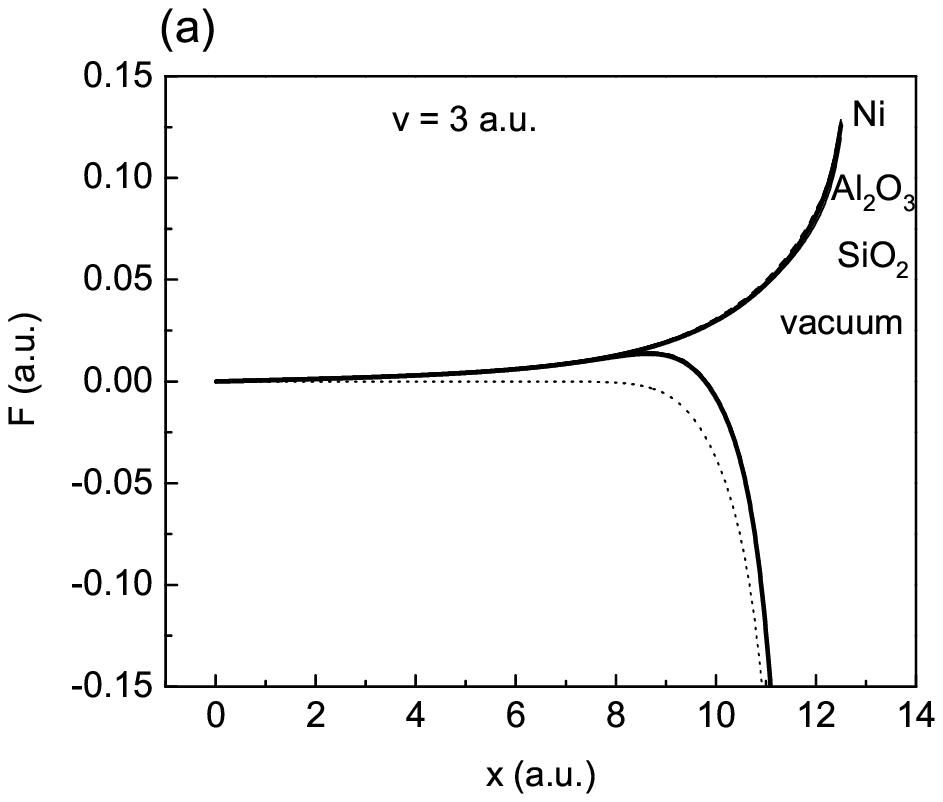}
\includegraphics[width=0.46\textwidth]{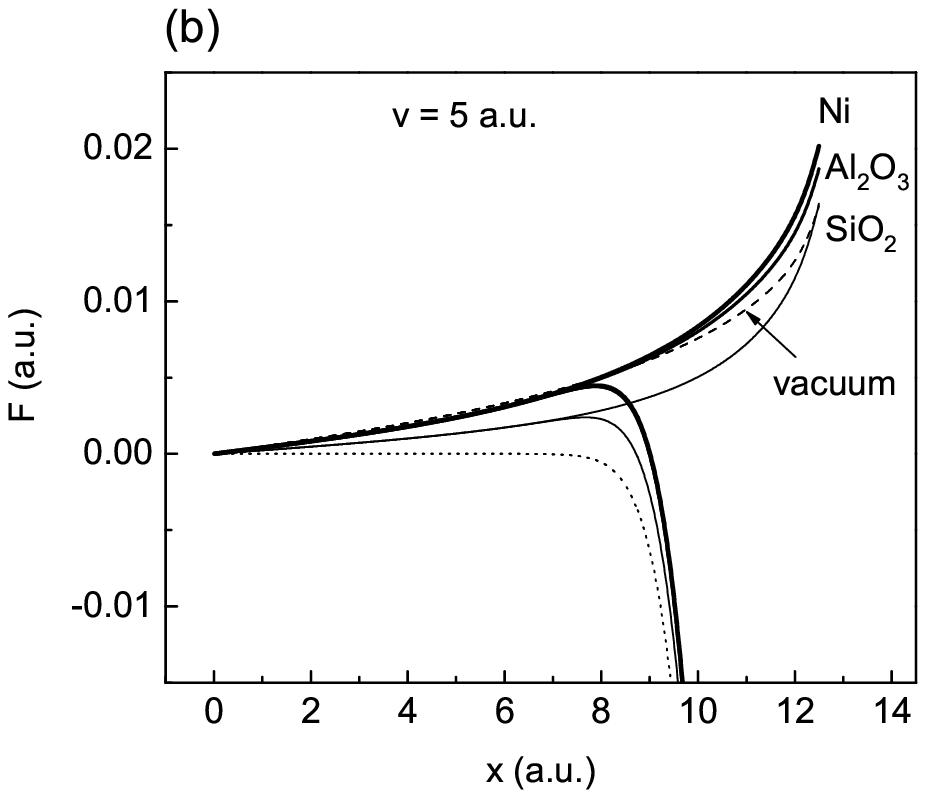}
\includegraphics[width=0.46\textwidth]{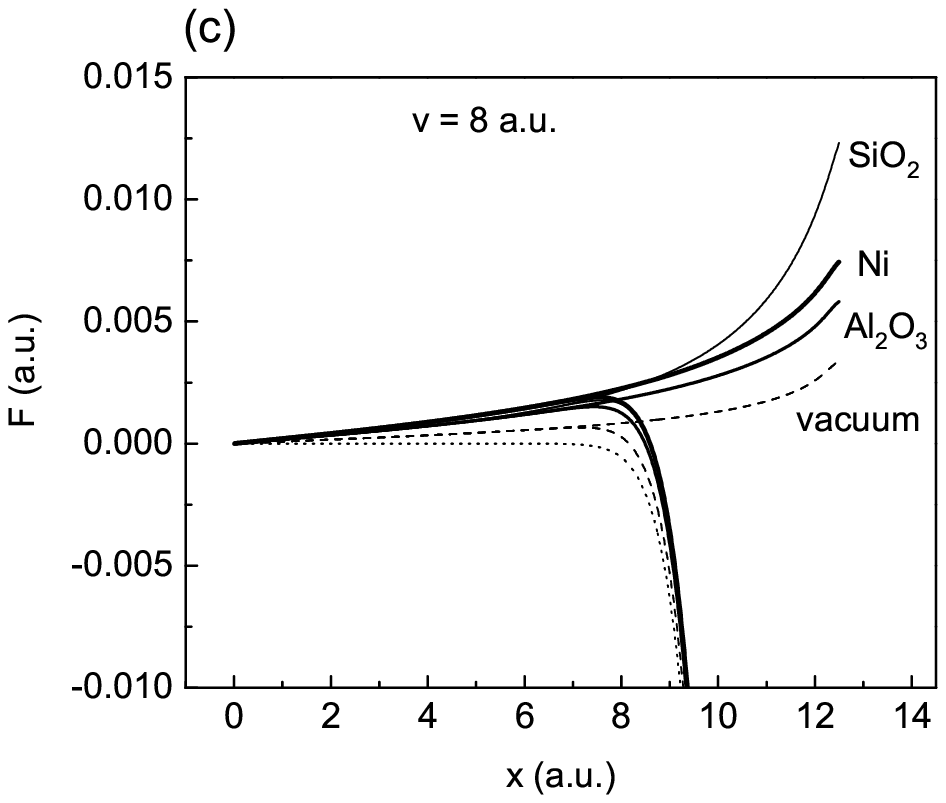}
\caption{The image force of a proton channeled in an (11,~9) SWNT in
vacuum (dashed curve) and encapsulated by SiO$_\mathrm{2}$,
Al$_\mathrm{2}$O$_\mathrm{3}$ and Ni channels (thin, medium, and
thick solid curves) as functions of the proton position $x$ in a.u.
across the nanotube radius, for three proton speeds: (a) $v$ = 3
a.u., (b) $v$ = 5 a.u. and (c) $v$ = 8 a.u. The dotted curve gives
the repulsive force, $F_\mathrm{rep}$, while the lower branches of
solid and dashed curves give the attractive image force alone,
$F_\mathrm{image}$, and the upper branches give the total force $F$,
due to an (11,~9) SWNT in vacuum (dashed curve) and encapsulated by
the three channels (solid curves). The radius of the nanotube is $a$
= 13.01 a.u. and the radius of the dielectric channel is $b$ = 16.22
a.u.}
\label{fig02}
\end{figure}

\begin{figure}
\centering
\includegraphics[width=0.48\textwidth]{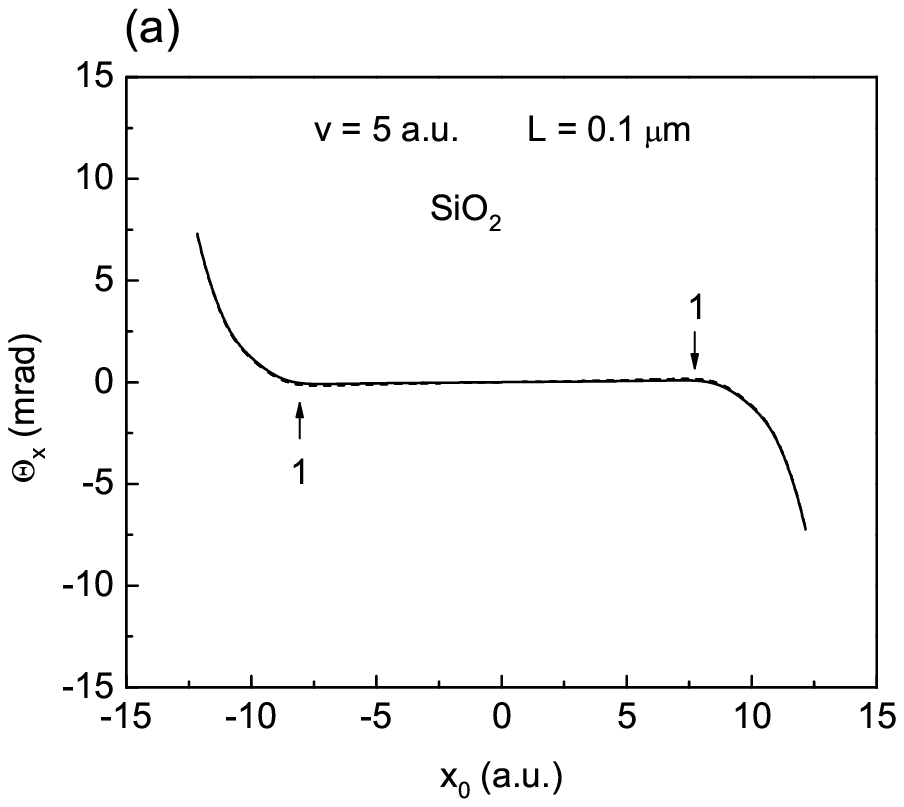}
\includegraphics[width=0.48\textwidth]{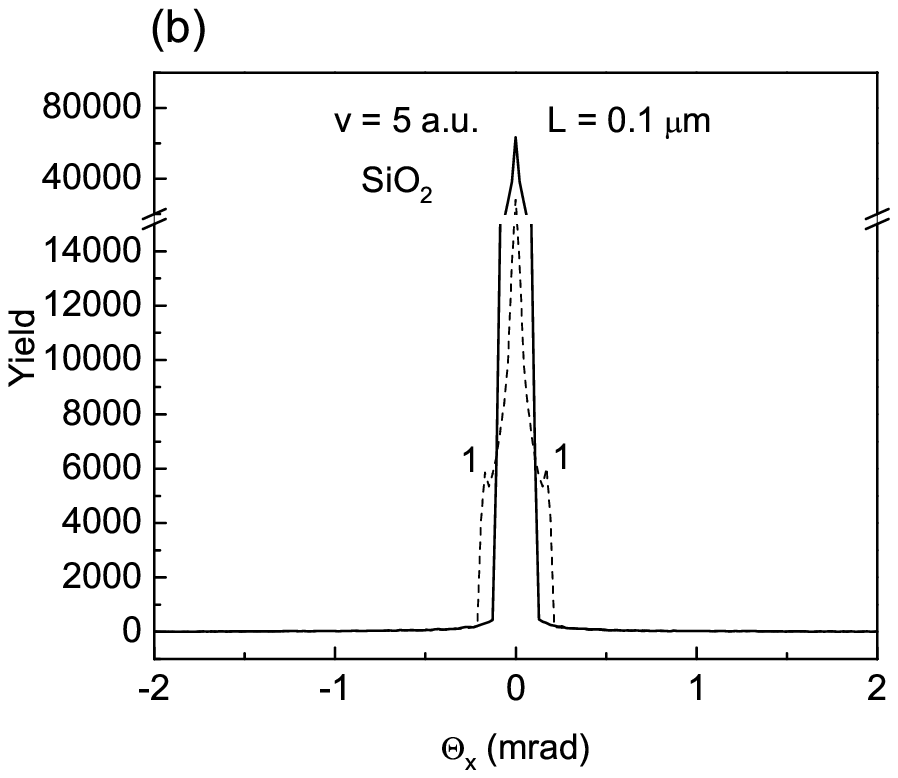}
\caption{The (a) deflection functions and (b) corresponding angular
distributions of protons channeled in an (11,~9) SWNT in vacuum
(dashed curve) and encapsulated by a SiO$_\mathrm{2}$ channel (solid
curve). The proton speed is $v$ = 5 a.u., the nanotube length is $L$
= 0.1 $\mu$m, the nanotube radius is $a$ = 13.01 a.u., and the
dielectric channel radius is $b$ = 16.22 a.u. The angular
distribution's bin size is 0.0213 mrad.}
\label{fig03}
\end{figure}

\begin{figure}
\centering
\includegraphics[width=0.48\textwidth]{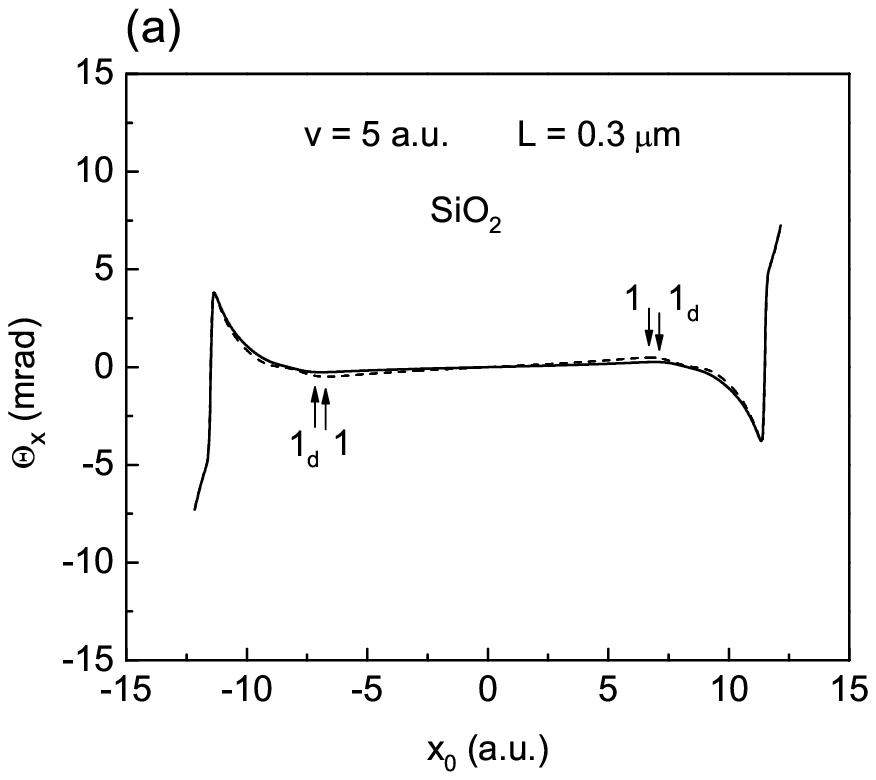}
\includegraphics[width=0.48\textwidth]{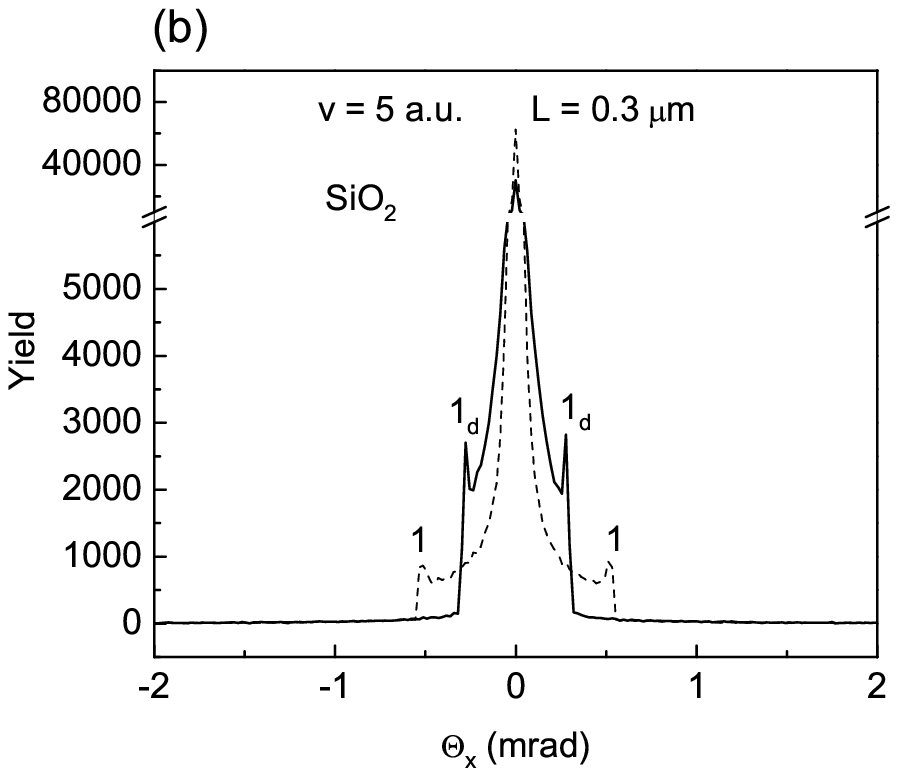}
\caption{The (a) deflection functions and (b) corresponding angular
distributions of protons channeled in an (11,~9) SWNT in vacuum
(dashed curve) and encapsulated by a SiO$_\mathrm{2}$ channel (solid
curve). The proton speed is $v$ = 5 a.u., the nanotube length is $L$
= 0.3 $\mu$m, the nanotube radius is $a$ = 13.01 a.u., and the
dielectric channel radius is $b$ = 16.22 a.u. The angular
distribution's bin size is 0.0213 mrad.}
\label{fig04}
\end{figure}

\begin{figure}
\centering
\includegraphics[width=0.48\textwidth]{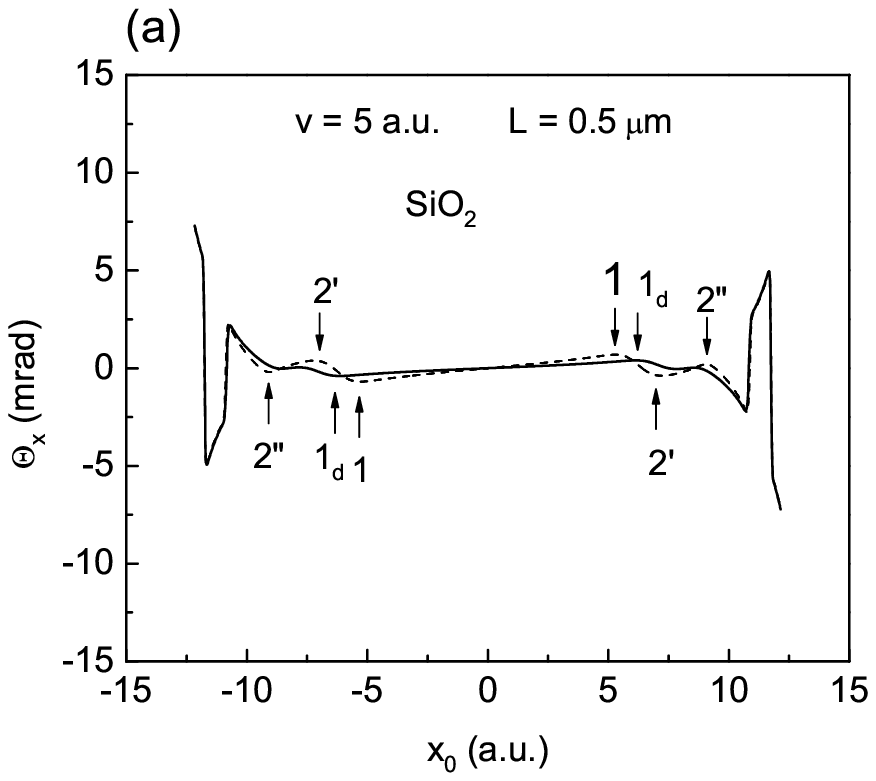}
\includegraphics[width=0.48\textwidth]{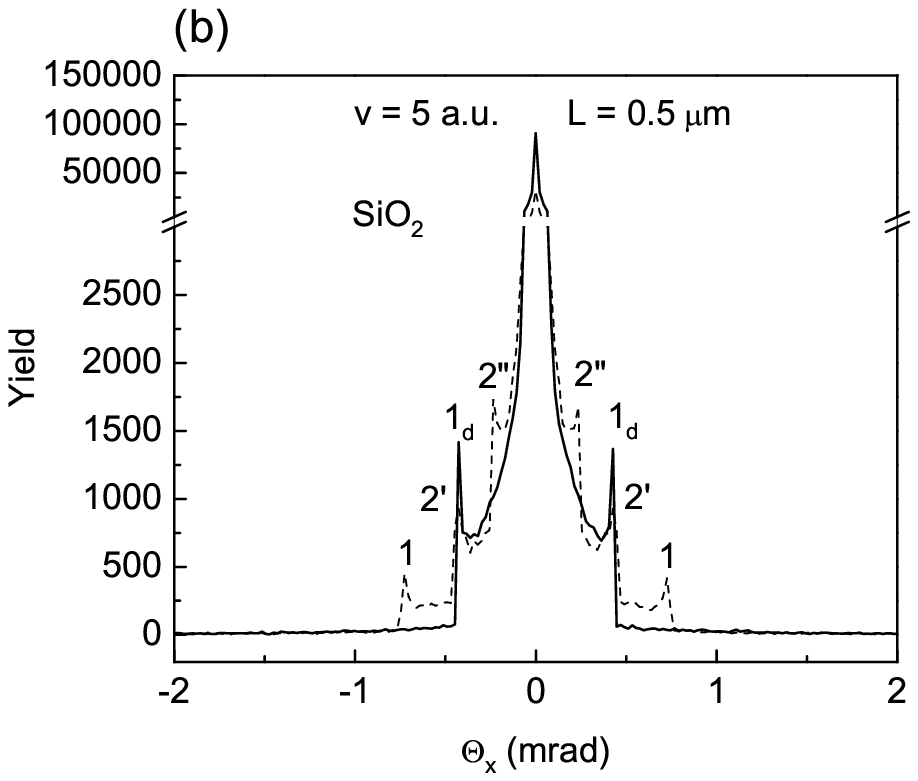}
\caption{The (a) deflection functions and (b) corresponding angular
distributions of protons channeled in an (11,~9) SWNT in vacuum
(dashed curve) and encapsulated by a SiO$_\mathrm{2}$ channel (solid
curve). The proton speed is $v$ = 5 a.u., the nanotube length is $L$
= 0.5 $\mu$m, the nanotube radius is $a$ = 13.01 a.u. and the
dielectric channel radius is $b$ = 16.22 a.u. The angular
distribution's bin size is 0.0213 mrad.}
\label{fig05}
\end{figure}

\begin{figure}
\centering
\includegraphics[width=0.48\textwidth]{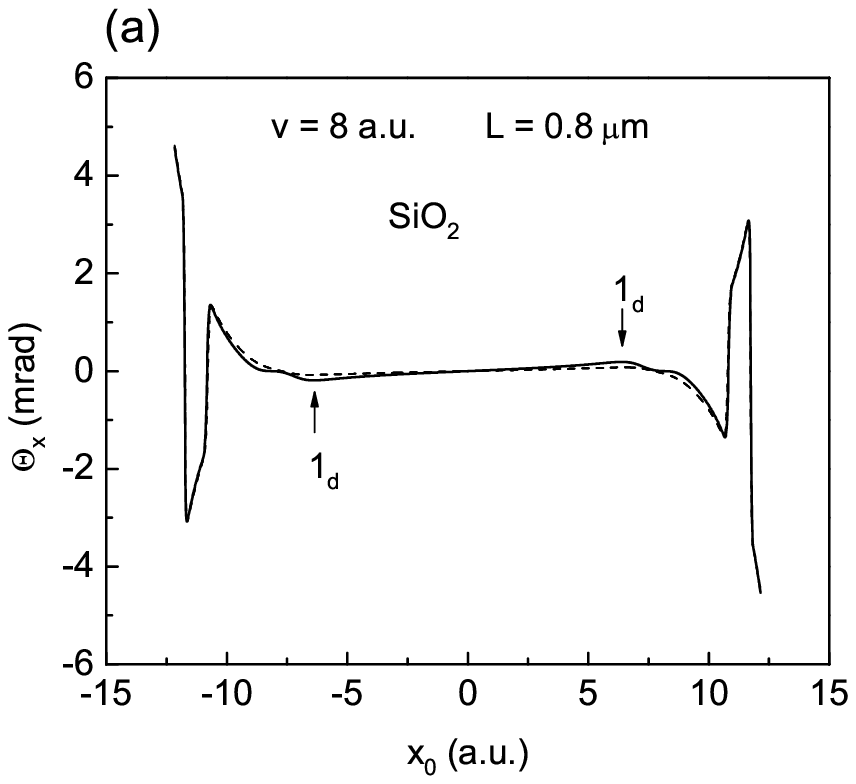}
\includegraphics[width=0.48\textwidth]{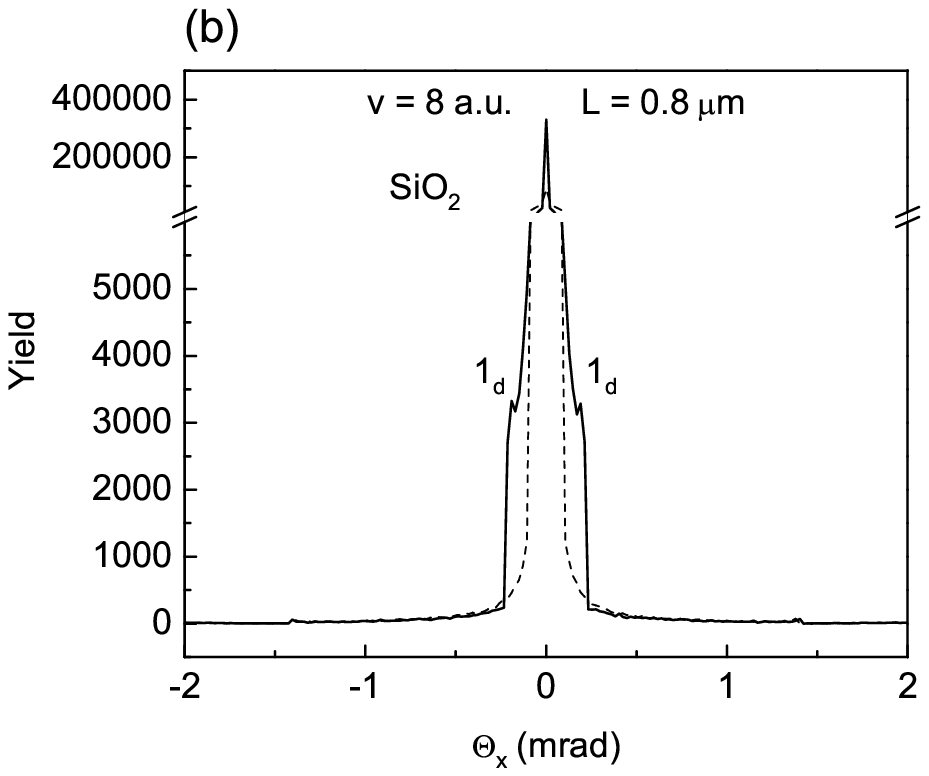}
\caption{The (a) deflection functions and (b) corresponding angular
distributions of protons channeled in an (11,~9) SWNT in vacuum
(dashed curve) and encapsulated by a SiO$_\mathrm{2}$ channel (solid
curve). The proton speed is $v$ = 8 a.u., the nanotube length is $L$
= 0.8 $\mu$m, the nanotube radius is $a$ = 13.01 a.u. and the
dielectric channel radius is $b$ = 16.22 a.u. The angular
distribution's bin size is 0.0213 mrad.}
\label{fig06}
\end{figure}

\begin{figure}
\centering
\includegraphics[width=0.48\textwidth]{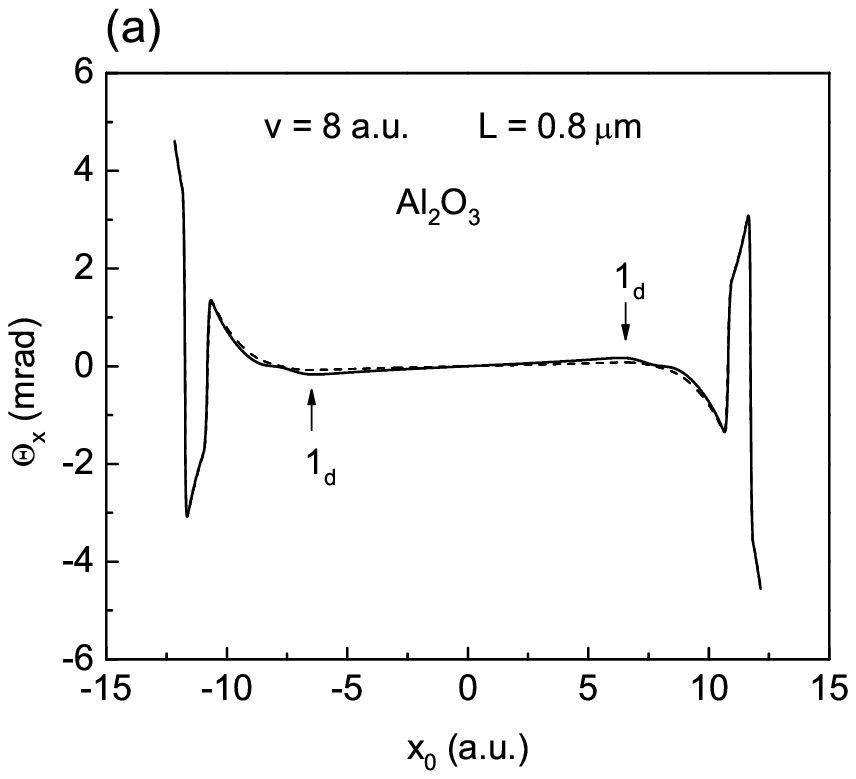}
\includegraphics[width=0.48\textwidth]{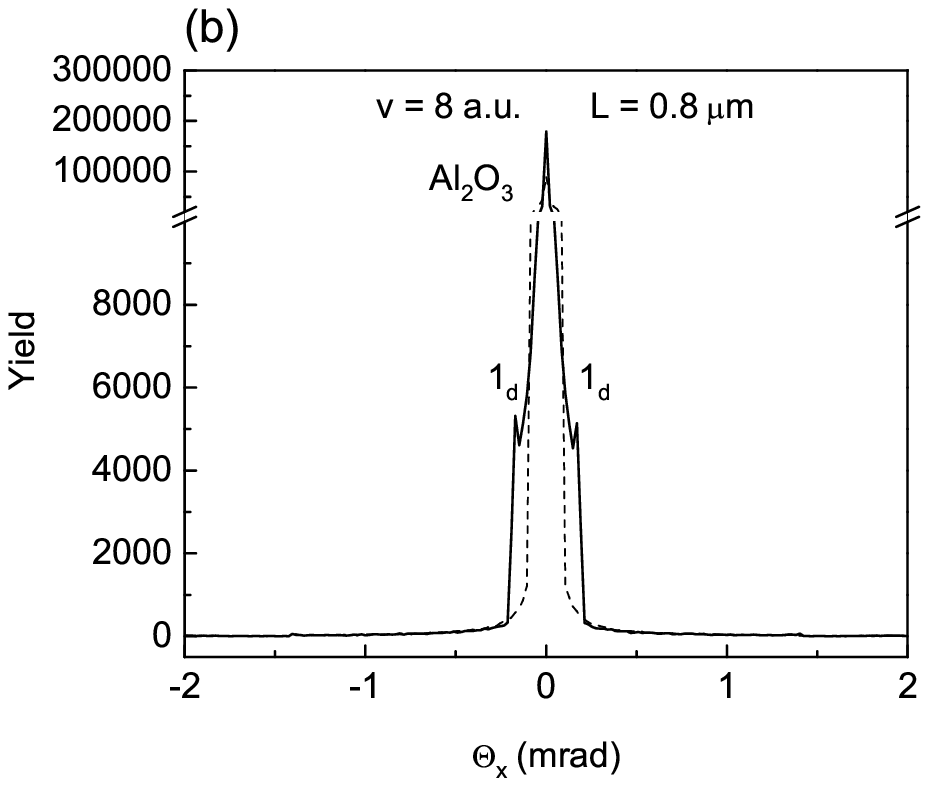}
\caption{The (a) deflection functions and (b) corresponding angular
distributions of protons channeled in an (11,~9) SWNT in vacuum
(dashed curve) and encapsulated by an Al$_\mathrm{2}$O$_\mathrm{3}$
channel (solid curve). The proton speed is $v$ = 8 a.u., the
nanotube length is $L$ = 0.8 $\mu$m, the nanotube radius is $a$ =
13.01 a.u. and the dielectric channel radius is $b$ = 16.22 a.u. The
angular distribution's bin size is 0.0213 mrad.}
\label{fig07}
\end{figure}

\begin{figure}
\centering
\includegraphics[width=0.48\textwidth]{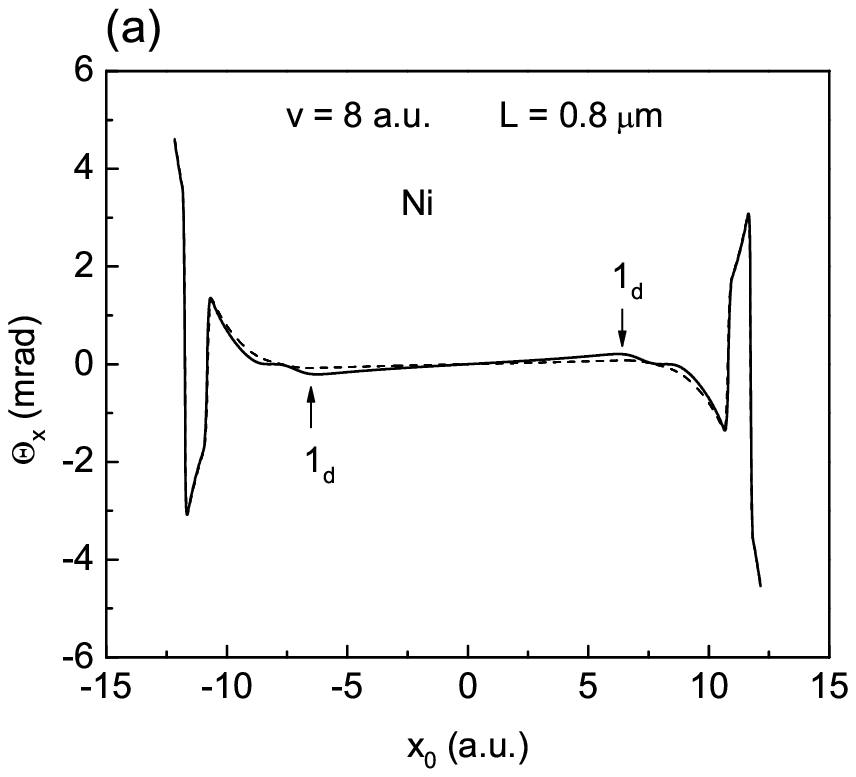}
\includegraphics[width=0.48\textwidth]{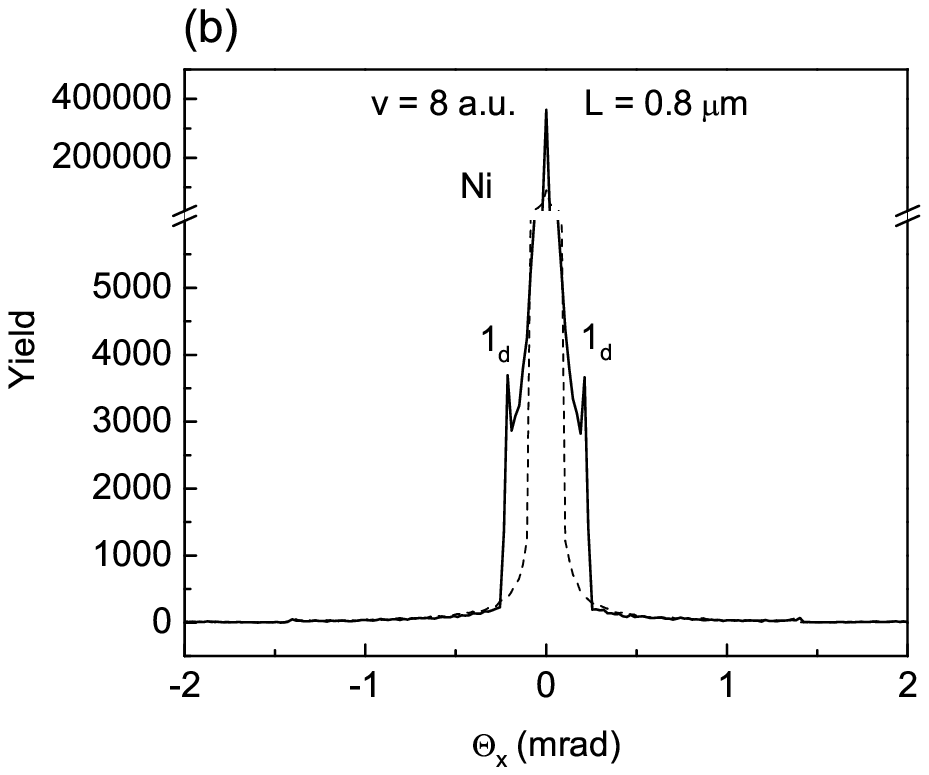}
\caption{The (a) deflection functions and (b) corresponding angular
distributions of protons channeled in an (11,~9) SWNT in vacuum
(dashed curve) and encapsulated by a Ni channel (solid curve). The
proton speed is $v$ = 8 a.u., the nanotube length is $L$ = 0.8
$\mu$m, the nanotube radius is $a$ = 13.01 a.u. and the dielectric
channel radius is $b$ = 16.22 a.u. The angular distribution's bin
size is 0.0213 mrad.}
\label{fig08}
\end{figure}

\begin{figure}
\centering
\includegraphics[width=0.48\textwidth]{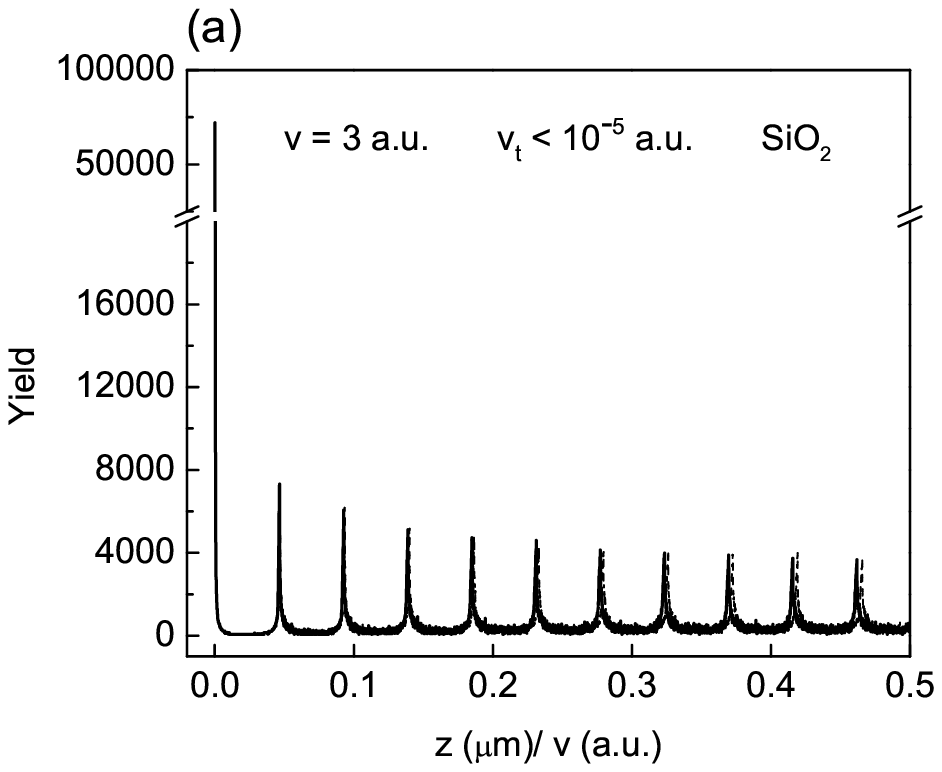}
\includegraphics[width=0.48\textwidth]{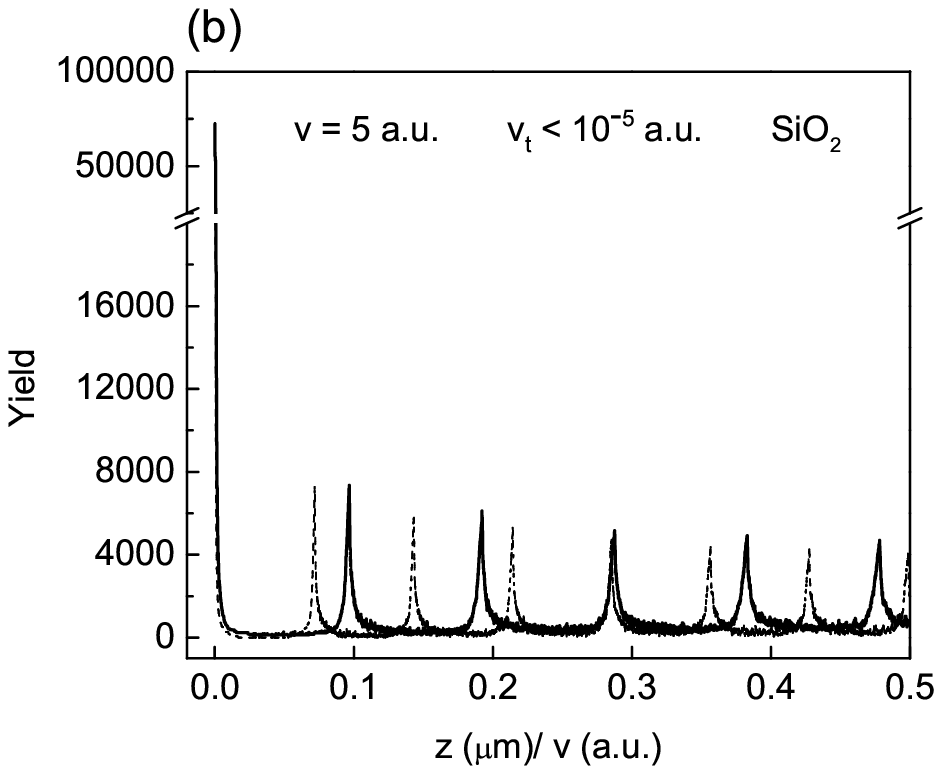}
\includegraphics[width=0.48\textwidth]{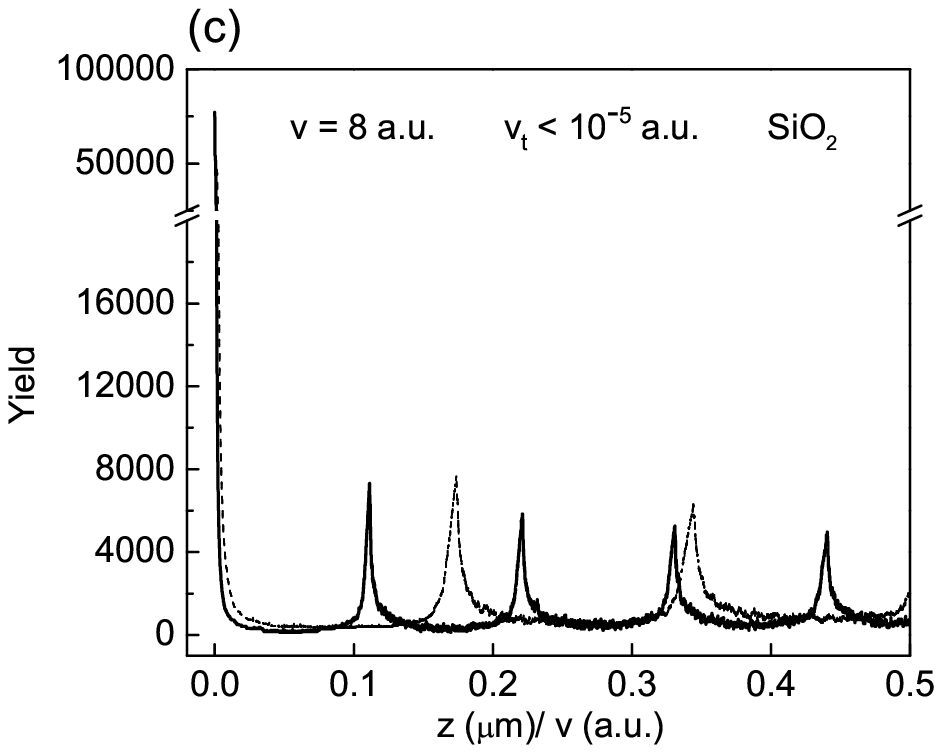}
\caption{Dependence of the zero-degree focusing (ZDF) versus dwell
time $L/v$ (in units of $\mu$m/$v_\mathrm{b}$ = 0.4571 ps) for
protons channeled in an (11,~9) SWNT in vacuum (dashed curve) and
encapsulated by a SiO$_\mathrm{2}$ channel (solid curve). The
longitudinal proton speeds are (a) $v$ = 3 a.u., (b) $v$ = 5 a.u.,
and (c) $v$ = 8 a.u., while the proton yield at ZDF is obtained by
collecting all transverse proton speeds such that $v_\mathrm{t} <
10^{-5}$ a.u.}
\label{fig09}
\end{figure}

\begin{figure}
\centering
\includegraphics[width=0.48\textwidth]{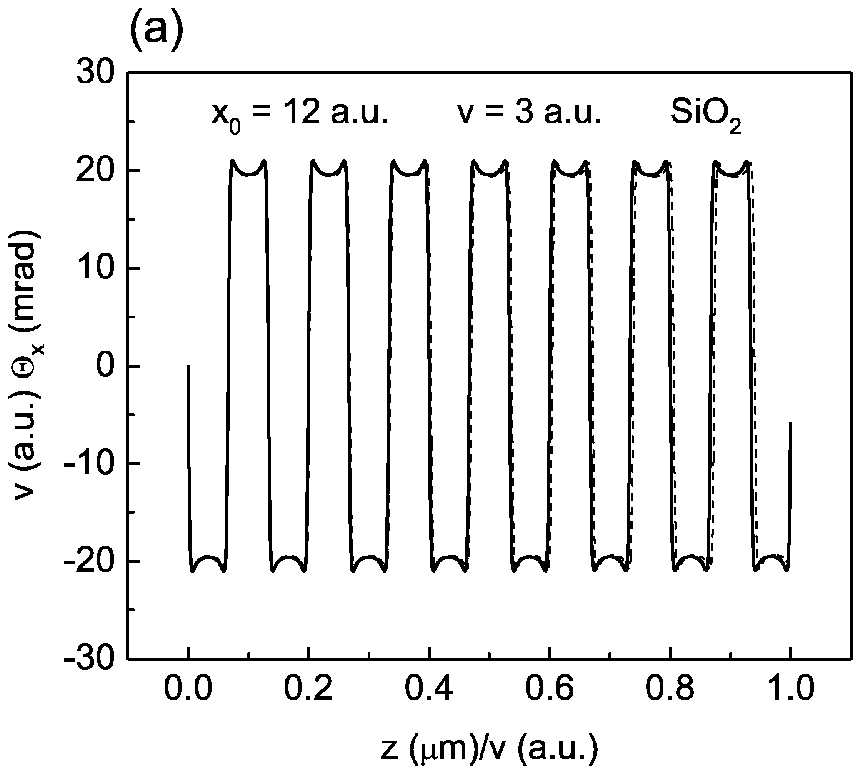}
\includegraphics[width=0.48\textwidth]{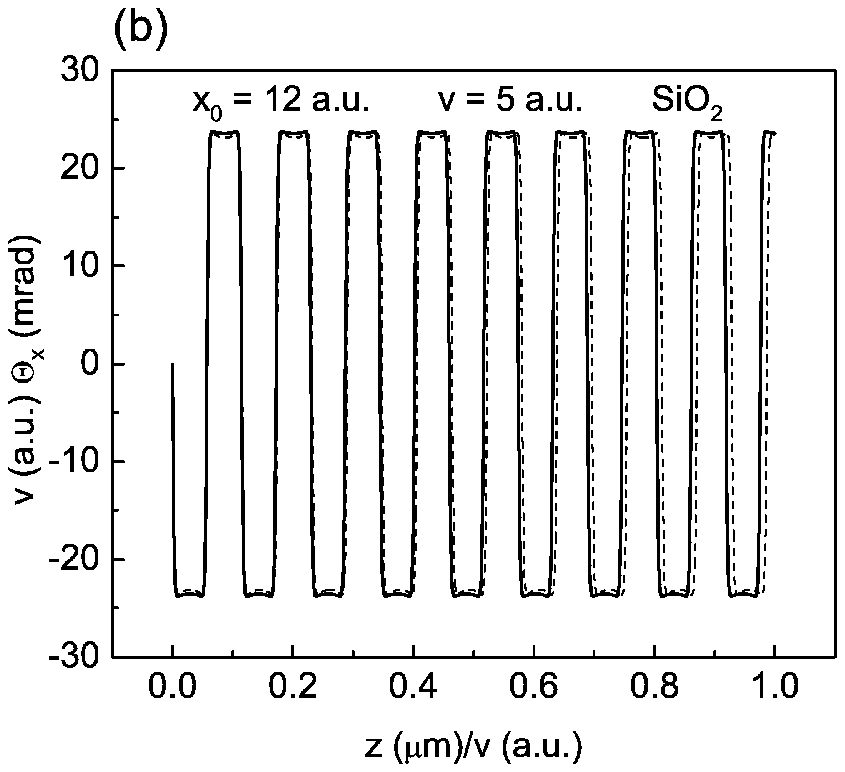}
\includegraphics[width=0.48\textwidth]{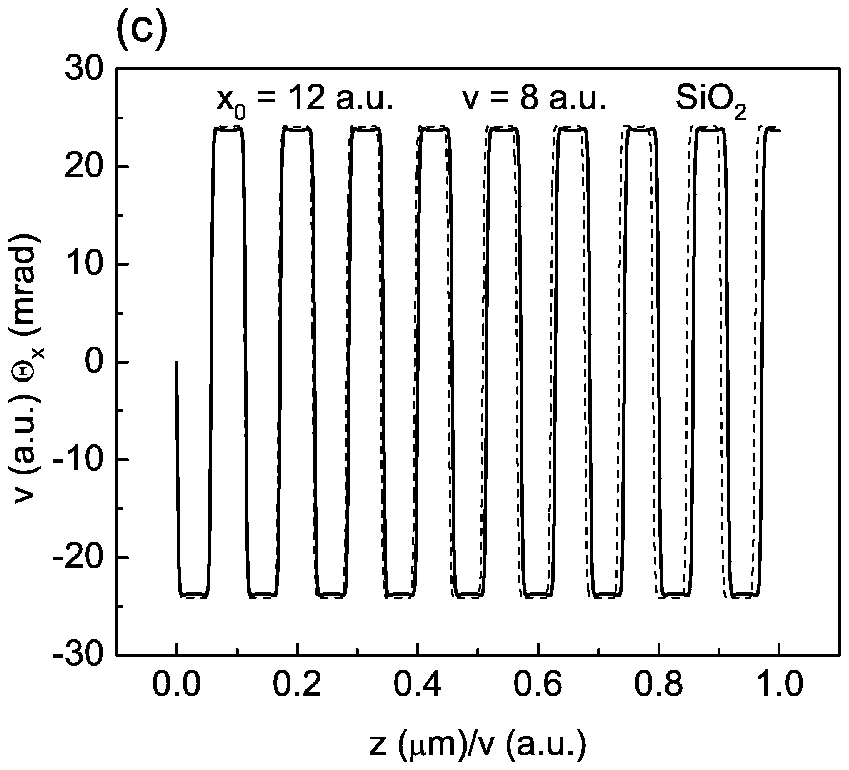}
\caption{Dependence of the proton deflection angle
$\Theta_\mathrm{x}$ in mrad multiplied by the proton speed $v$ in
a.u. on the dwell time $L/v$ (in units of $\mu$m/$v_\mathrm{b}$ =
0.4571 ps) for a proton impact parameter of $x_\mathrm{0}$ = 12 a.u.
and the three protons speeds (a) $v$ = 3 a.u., (b) $v$ = 5 a.u. and
(c) $v$ = 8 a.u., due to an (11,~9) SWNT in vacuum (dashed curve)
and encapsulated by SiO$_\mathrm{2}$ channels (solid curve).}
\label{fig10}
\end{figure}

\begin{figure}
\centering
\includegraphics[width=0.48\textwidth]{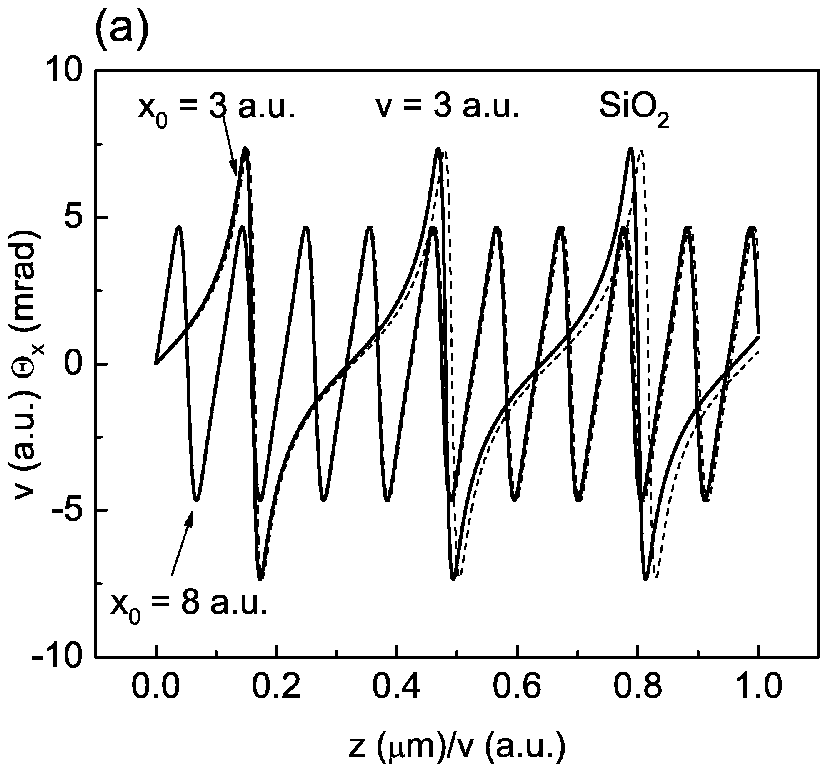}
\includegraphics[width=0.48\textwidth]{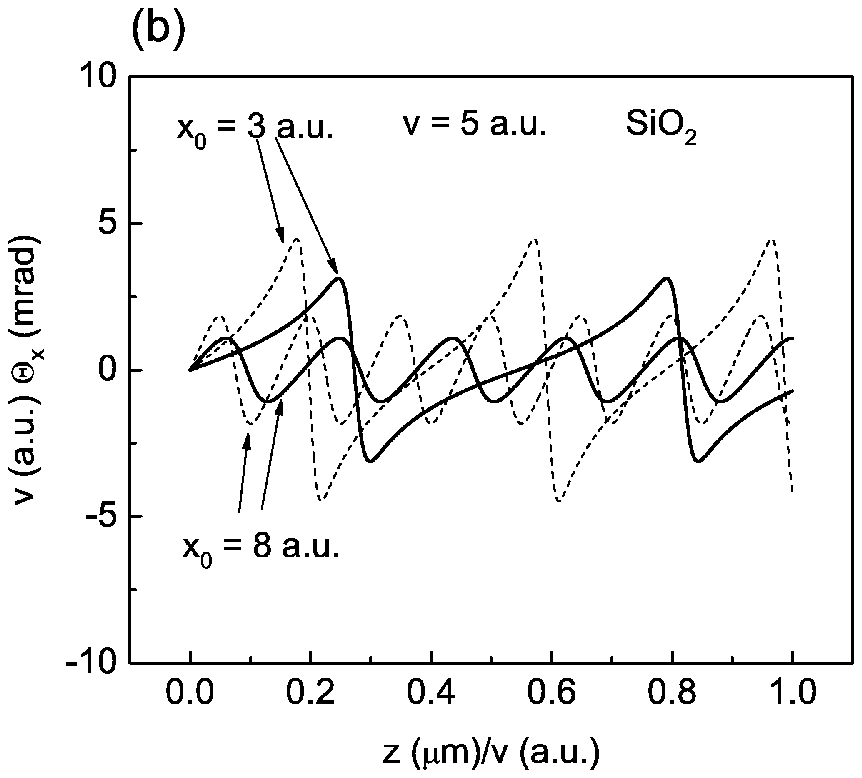}
\includegraphics[width=0.48\textwidth]{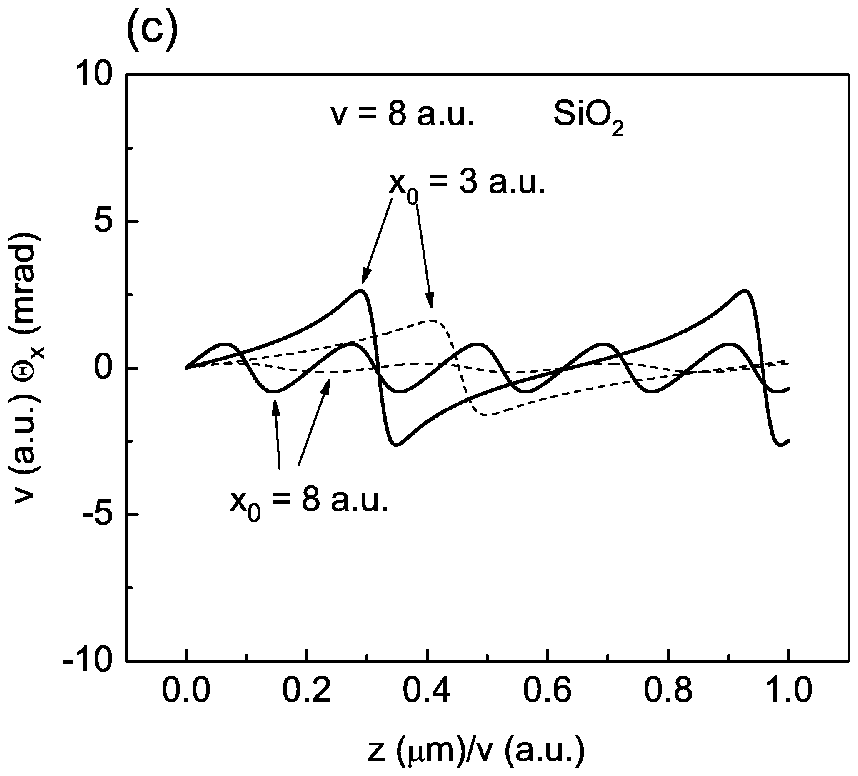}
\caption{Dependence of the proton deflection angle
$\Theta_\mathrm{x}$ in mrad multiplied by the proton speed $v$ in
a.u. on the dwell time $L/v$ (in units of $\mu$m/$v_\mathrm{b}$ =
0.4571 ps) for proton impact parameters of $x_\mathrm{0}$ = 3 a.u.
and $x_\mathrm{0}$ = 8 a.u. and three proton speeds (a) $v$ = 3
a.u., (b) $v$ = 5 a.u. and (c) $v$ = 8 a.u., due to an (11,~9) SWNT
in vacuum (dashed curve) and encapsulated by SiO$_\mathrm{2}$
channels (solid curve).}
\label{fig11}
\end{figure}

In this section we shall first discuss in Figures \ref{fig01} and
\ref{fig02} respectively the image force and the total interaction
force for a proton channeled in an (11,~9) SWNT in vacuum and
encapsulated by SiO$_\mathrm{2}$, Al$_\mathrm{2}$O$_\mathrm{3}$ and
Ni channels. Subsequently, we shall analyze the dynamic polarization
effects on the angular distributions of protons after channeling at
a speed of $v$ = 5 a.u. through an (11,~9) nanotube placed in vacuum
and in SiO$_\mathrm{2}$ for three nanotube lengths, $L$ = 0.1, 0.3,
and 0.5 $\mu$m in Figures \ref{fig03}, \ref{fig04} and \ref{fig05},
respectively. This analysis will be followed by a comparison of the
dynamic polarization effects for different proton speeds but with a
fixed dwell time, $T$ = $L/v$ = 0.1 $\mu$m/$v_\mathrm{B}$ = 0.04571
ps (with $v_\mathrm{B}$ being the Bohr speed), for nanotubes in
vacuum and in SiO$_\mathrm{2}$. Next, we shall compare the effects
of different media, namely SiO$_\mathrm{2}$,
Al$_\mathrm{2}$O$_\mathrm{3}$, and Ni, for the case of $L$ = 0.8
$\mu$m, $v$ = 8 a.u., as shown in Figures \ref{fig06}, \ref{fig07}
and \ref{fig08}, respectively. Finally, we shall comment on the zero
degree focusing (ZDF) effect for proton speeds $v$ = 3 a.u., $v$ = 5
a.u. and $v$ = 8 a.u. for nanotubes of various lengths, placed in
vacuum and in SiO$_\mathrm{2}$, as shown in Figure \ref{fig09}.
Qualitative analysis of the results will also be presented in terms
of typical proton trajectories in the nanotube in vacuum and in
SiO$_\mathrm{2}$, shown in Figures \ref{fig10} and \ref{fig11}.

Figure \ref{fig01} gives the image force of a proton traveling
paraxially at $r$ = 3 (lower), 7 (middle) and 10 a.u. (upper), with
speed $v$, due to an (11,~9) SWNT in vacuum (dashed curves) and
encapsulated by three dielectric materials (solid curves). The
radius of this nanotube is $a \approx$ 13.01 a.u. while the radius
of the dielectric channel is $b \approx a$ + 3.21 a.u. $\approx$
16.22 a.u., where we have used carbon's van der Waals radius
($\approx$ 3.21 a.u.) to approximate the separation between the
nanotube and the nearby dielectric surface. In the case a nickel
channel surrounding the nanotube, DFT based structural minimization
calculations performed using SIESTA \cite{sole02} for graphene on a
Ni(111) surface also yielded a separation of approximately 0.17 nm.
In Fig. \ref{fig01}(a) we describe the surrounding silicon dioxide
(SiO$_\mathrm{2}$) by a dielectric constant of 3.9 \cite{swar00}, in
Fig. \ref{fig01}(b) we model the dielectric response of anodic
aluminum oxide (Al$_\mathrm{2}$O$_\mathrm{3}$) following the method
described in Ref. \onlinecite{aris01aris01}, and in Fig.
\ref{fig01}(c) we model the dielectric response of Ni metal
following the method described in Ref. \onlinecite{kwei93}. The
solid and dashed curves in Figs. \ref{fig01} indicate that, for
proton speeds below $v \approx$ 3 a.u., the dielectrics have little
influence on the image force, for proton speeds between $v \approx$
3 a.u. and $v \approx$ 6 a.u. the image force is somewhat smaller in
the presence of dielectrics, while, for speeds above $v \approx$ 6
a.u. the image force is increased compared to the vacuum case. These
results may be explained by the plasmon hybridization in the
nanotube-dielectric system, as previously described in Ref.
\onlinecite{mowb06}. In brief, protons moving at speeds below $v
\approx$ 3 a.u. do not induce plasma oscillations in the nanotube,
so that the dielectric media are completely screened from the ion
and do not influence the ion's self energy. Above this speed plasma
oscillations appear, giving incomplete screening of the dielectric
media by the nanotube via strong plasmon hybridization. As a result,
dielectric media strongly affect the proton's self energy and the
resulting image force at intermediate proton speeds. At high speeds,
the nanotube becomes increasingly transparent to the proton, so that
the dynamic-polarization effects tend to be dominated by the
surrounding dielectric media. This is further elaborated in Fig.
1(c), where we also show the image force on proton at the same three
distances inside Ni channels containing \emph{no} encapsulated
nanotube. Specifically, we display the results for a channel of
radius $b$ = 16.22 a.u. (dotted lines) showing how dynamic
polarization of nickel becomes prominent in the combined
nanotube-channel system only at high speeds. For the sake of
comparing different channeling systems, we also show in Fig. 1(c)
the results for image force in a Ni channel with its radius equal to
that of a (11,~9) nanotube, $b = a$ = 13.01 a.u. (thin solid lines).
One can see that the image forces inside the nanotube in vacuum and
in a Ni channel of the same radius have generally comparable
magnitudes, with the case of a Ni channel undergoing stronger
polarization at high proton speeds, say $v >$ 6 a.u., and the
nanotube providing a stronger image force at the intermediate to low
proton speeds, say $v \propto$ 2-5 a.u.

We further analyze in Fig. \ref{fig02} the total force on proton
channeling, $F$ = $F_\mathrm{rep}$ + $F_\mathrm{image}$ as a
function of the position $x$ (in a.u.) across the nanotube radius,
for proton speeds of (a) $v$ = 3 a.u., (b) $v$ = 5 a.u., and (c) $v$
= 8 a.u. Here, the dashed curves denote the case of an (11,~9)
nanotube in vacuum, while the solid curves of various thicknesses
denote the effects of the three surrounding media, SiO$_\mathrm{2}$,
Al$_\mathrm{2}$O$_\mathrm{3}$ and Ni, by thin, medium, and thick
solid curves, respectively. The lower branches of these curves show
the behavior of the image force $F_\mathrm{image}$ alone close to
the nanotube, while the upper branches show the total force $F$ as a
result of adding the bare repulsive force $F_\mathrm{rep}$ (shown by
dotted lines) to the image force. Although close to the nanotube the
total force is obviously dominated by the repulsive force due to the
carbon atoms, the effects of the dielectric media on the attractive
interactions due to the image force are greatly affected by both the
proton speed and the type of the surrounding material. Figure
\ref{fig02}(a) shows practically no effects of the dielectric media
for a proton speed of $v$ = 3 a.u., as expected from Fig.
\ref{fig01}. At an intermediate proton speed of $v$ = 5 a.u, while
the curves in Fig. \ref{fig02}(b) for Al$_\mathrm{2}$O$_\mathrm{3}$,
Ni and vacuum are very close to each other, especially in the inner
part of the nanotube, the case of SiO$\mathrm{2}$ displays a
surprisingly weaker image force almost up to the nanotube wall. On
the other hand, for $v$ = 8 a.u., the image forces shown in Fig.
\ref{fig02}(c) have similar values for all three dielectric media,
especially in the inner part of the nanotube, whereas the image
force for the case of nanotube in vacuum is substantially weaker at
all distances. As commented in reference to Fig. \ref{fig01}, the
nanotube becomes increasingly transparent at high proton speeds, so
that any dynamic-polarization effects will be due to the
polarization of the surrounding dielectrics \cite{mowb06}. This
implies that, while the image interaction will be heavily suppressed
at high proton speeds for the nanotube in vacuum, it can remain
operational in the presence of surrounding media and, in fact, will
be dominated by their dielectric properties as if carbon nanotube
was not there. Thus, since the image force is crucial for the
appearance of rainbows and the ZDF in ion angular distributions
after channeling through short carbon nanotubes, we expect strong
effects of the surrounding media, especially at higher (but still
non-relativistic) speeds.

In Figures \ref{fig03}-\ref{fig05}, we show the results of proton
channeling through the nanotube in vacuum (dashed curves) and
encapsulated by a SiO$_\mathrm{2}$ channel (solid curves), by both
the deflection function $\Theta_\mathrm{x}(x_\mathrm{0})$ in panels
(a), and the corresponding angular distributions in panels (b). In
Fig. \ref{fig03}, where the proton speed is $v$ = 5 a.u. and the
nanotube length is $L$ = 0.1 $\mu$m, we find a pair of very shallow
extrema, labeled 1, of the dashed curve in panel (a), whereas the
solid curve does not exhibit such extrema. As a consequence, the
corresponding angular distributions in Fig. \ref{fig03}(b) display,
besides massive central peaks, also two very small rainbow peaks,
also labeled 1, in the case when the nanotube is in vacuum (dashed
curve). However, we find no rainbow peaks when the nanotube is
surrounded by SiO$_\mathrm{2}$ (solid curve). On the other hand, the
yield in the central maximum for the SiO$_\mathrm{2}$ case is found
to be almost three times larger than in the case of a nanotube in
vacuum. These findings can be explained by examining the results in
Fig. \ref{fig02}(b) shown by the thin solid curve for the
SiO$_\mathrm{2}$ case in comparison to those shown by the dashed
curve for a nanotube in vacuum. The rainbow effect is missing in the
case of SiO$_\mathrm{2}$ because the image force is too weak
compared to the vacuum case, and the protons are not pulled far
enough from their initial direction in such a short nanutobe. On the
other hand, the image potential well (not shown here) is found to be
shallower and broader in the case of SiO$_\mathrm{2}$ implying that
a larger fraction of the incident protons will be very near the
initial direction, giving more flux of undeflected particles than in
the vacuum case. Increasing the nanotube length to $L$ = 0.3 $\mu$m,
as shown in Fig. \ref{fig04}, amounts to a shift of the extrema 1 in
the vacuum case (dashed curves), and also to the appearance of  a
pair of extrema, labeled by 1$_\mathrm{d}$, in the solid curves due
to the image force when the nanotube is in a dielectric medium
(SiO$_\mathrm{2}$). A further increase of the nanotube length to $L$
= 0.5 $\mu$m, as shown in Fig. \ref{fig05}, gives rise to multiple
rainbows when the carbon nanotube is in vacuum. As discussed in Ref.
\onlinecite{borka06}, the rainbows labeled by 1 belong to the first
class (rainbow trajectories which have experienced one deflection
within the image-generated potential well), while the rainbows
labeled by 2$'$ and 2$''$ belong to the second class (rainbow
trajectories which have experienced two deflections within the
potential well). However, no new rainbow peaks appear in the solid
curves shown in Fig. \ref{fig05} when the nanotube is in
SiO$_\mathrm{2}$, besides the pair of peaks labeled by
1$_\mathrm{d}$, which are now shifted in comparison to those in Fig.
\ref{fig04}. This inability of the image force to create multiple
rainbows in the case of the surrounding SiO$_\mathrm{2}$ channel can
again be explained by comparing the thin solid line for
SiO$_\mathrm{2}$ with the dashed line for a nanotube in vacuum,
shown in Fig. \ref{fig02}(b).

We further concentrate on the velocity dependencies in the effects
of dielectrics on the image force, while eliminating the cumulative
effects on ion deflection due to the increasing nanotube length.
This is done by looking into various cases with fixed dwell time,
$T$ = $L/v$ = 0.1 $\mu$m/$v_\mathrm{B}$, for nanotubes in vacuum and
in SiO$_\mathrm{2}$. Considering low proton speeds, say below $v
\approx$ 3 a.u., we note that the results for all three surrounding
media (SiO$_\mathrm{2}$, Al$_\mathrm{2}$O$_\mathrm{3}$, or Ni)
should be virtually identical to the results for an (11,~9) nanotube
in vacuum. This is because protons at such speeds do not induce
plasma oscillations in the carbon nanotube, so that the influence of
the dielectric media is completely screened, as shown in Figs.
\ref{fig01} and \ref{fig02}(a). The results for the combination of
parameters $L$ = 0.3 $\mu$m, $v$ = 3 a.u. are displayed in Figure 4
of Ref. \onlinecite{borka06}, where one can see five pairs of the
extremal points, labeled by 1, 2$'$, 2$''$, 3$'$ and 3$''$, which
result from one (1), two (2$'$ and 2$''$) and three (3$'$ and 3$''$)
deflections of the rainbow trajectories within the image-generated
potential well. For the combination of parameters $L$ = 0.5 $\mu$m,
$v$ = 5 a.u., as shown in Fig. \ref{fig05}, one notices a
significant depletion of the number of rainbow peaks, down to three
pairs (1, 2$'$ and 2$''$) for the nanotube in vacuum, and only one
pair (1$_\mathrm{d}$) for surrounding SiO$_\mathrm{2}$. The further
reduction of the number of rainbow peaks with increasing proton
speed is illustrated in Fig. \ref{fig06} for the combination of
parameters $L$ = 0.8 $\mu$m, $v$ = 8 a.u. Here, we notice that the
massive central peak is wider than in Fig. \ref{fig05}, while the
rainbow effect has completely disappeared for the nanotube in
vacuum. This is due to the diminished image force on protons at such
a high speed. On the other hand, the one rainbow peak
(1$_\mathrm{d}$) from Fig. \ref{fig05} has remained in Fig.
\ref{fig06} for the case of a nanotube surrounded by
SiO$_\mathrm{2}$, although this peak is now very small. This
persistence of the rainbow peak for a surrounding dielectric can be
explained by the "transparency" of nanotubes at high proton speeds
\cite{mowb06}, where the image force is dominated by the
polarization of the surrounding dielectric, as shown in Fig.
\ref{fig02}(c).

We also compare the case of proton speed $v$ = 8 a.u. and nanotube
length $L$ = 0.8 $\mu$m for the nanotube in SiO$_\mathrm{2}$,
Al$_\mathrm{2}$O$_\mathrm{3}$ and Ni, as shown in Figures
\ref{fig06}, \ref{fig07} and \ref{fig08}, respectively. One notices
that the positions of the rainbow extrema, labeled by 1$\mathrm{d}$,
hardly change for the solid curves shown in Figs.
\ref{fig06}-\ref{fig08} for different dielectrics. This can be
explained by the relatively close proximity of the three image-force
curves for nanotubes in dielectrics, shown in Fig. \ref{fig02}(c)
for all three dielectrics, at distances where extremal points occur
in the corresponding deflection curves, as shown in Figs.
\ref{fig06}-\ref{fig08}.

As evidenced in Figs. \ref{fig03}-\ref{fig08}, the effects of
dielectric media on the dynamic polarization of carbon nanotubes
also affect quite strongly the central peaks in the angular
distributions of channeled ions. We therefore analyze next how these
effects change the ZDF in the case an (11,~9) nanotube in vacuum and
in a SiO$_\mathrm{2}$ channel. We first note that our simulations of
proton channeling through carbon nanotubes without the image force,
when proton trajectories are governed only by the repulsive
Doyle-Turner potential, did not yield any periodic peaking of ion
directions parallel to the nanotube. This is explained by the short
range of the Doyle-Turner potential, which is very steep near the
nanotube walls, so that protons with different impact parameters
undergo transversal oscillations with a wide range of periods.
Therefore, there is no single frequency of such oscillations that
may give rise to the periodic peaking of ion directions parallel to
the nanotube when their length increases. When the image potential
is included in our simulations, one finds parabolic regions in the
total potential near the minima. This allows for a broader range of
proton impact parameters that would give oscillations in the
transversal directions at almost the same frequency. When the
trajectories of such protons become almost parallel to the entrance
beam, one finds the effect of ZDF. In our simulations, we use as a
criterion for the ZDF that the proton speeds in the transverse
directions are $v_\mathrm{t} < 10^{-5}$ a.u.

Figure \ref{fig09} illustrates the effects of a surrounding medium
on the ZDF by showing the dependence of proton yield (for
$v_\mathrm{t} < 10^{-5}$ a.u.) on the dwell time, $L/v$, for protons
channeled in an (11,~9) SWNT in vacuum and encapsulated by a
SiO$_\mathrm{2}$ channel, where the proton speeds are (a) $v$ = 3
a.u., (b) $v$ = 5 a.u. and (c) $v$ = 8 a.u. While the surrounding
dielectric does not effect the ZDF for a proton speed of $v$ = 3
a.u., we notice strong effects on the ZDF for proton speeds $v$ = 5
a.u. and $v$ = 8 a.u. when compared to the case of the nanotube in
vacuum. Specifically, the period of the ZDF occurrence is seen to
increase steadily with increasing proton speed in the latter case,
owing to the weakening image interaction with the nanotubes in
vacuum, whereas this increase seems to saturate at higher proton
speeds for nanotubes in a SiO$_\mathrm{2}$ channel. Realizing that
the ZDF peaks appear more frequently when the image force is
stronger, one can easily refer to Figs. \ref{fig02} (b) and (c) to
explain why the first ZDF peaks for the nanotubes in
SiO$_\mathrm{2}$ appear in Figs. \ref{fig09} (b) and (c),
respectively, after and before the first peaks for the nanotubes in
vacuum.

As regards the yields in the central maxima on the (b) panels of
Figs. \ref{fig04}-\ref{fig06}, we have seen for the case of $v$ = 5
a.u. and $L$ = 0.3 $\mu$m, shown in Fig. \ref{fig04}(b), that the
yield when the nanotube is in vacuum is about twice the yield when
the nanotube is in SiO$_\mathrm{2}$. This is consistent with the
fact that, in Fig. \ref{fig09}(b), we are very close to the first
ZDF peak in vacuum for this proton speed and nanotube length. On the
other hand, in the case $v$ = 5 a.u. and $L$ = 0.5 $\mu$m, shown in
Fig. \ref{fig05}(b), and in the case $v$ = 8 a.u. and $L$ =0.8
$\mu$m, shown in Fig. \ref{fig06}(b), we notice that the yields of
the central maxima when the nanotubes are in SiO$_\mathrm{2}$ are
about three times the yields for the nanotubes in vacuum. This can
now be explained by noticing that, for these combinations of proton
speed and nanotube length, the yields are very close to the first
ZDF peaks in, respectively, Figs. \ref{fig09} (b) and (c) for
nanotubes in SiO$_\mathrm{2}$.

The results presented so far may be further elucidated by
considering the effects of the surrounding medium on typical proton
trajectories in carbon nanotubes, as displayed in Figs. \ref{fig10}
and \ref{fig11}. We note that, in the case without image potential,
there is only one type of proton trajectory, arising from proton
oscillations between the opposite sides of the nanotube wall. In
this case, the angular distributions of channeled protons through
different (11,~9) nanotubes are similar for the same dwell times
$L/v$, and their widths obey the law $v \cdot \Theta_\mathrm{w}$ =
constant for different protons speeds. When the image potential is
included, there are two characteristic types of proton trajectories,
arising from proton oscillations between the opposite sides of the
nanotube wall, and proton oscillations in the potential minima
generated by the image potential.

Figure \ref{fig10} shows dependence of the proton deflection angle
$\Theta_\mathrm{x}$ multiplied by the proton speed $v$ on the dwell
time $L/v$ for an impact parameter of $x_\mathrm{0}$ = 12 a.u. and
three different proton speeds (a) $v$ = 3 a.u., (b) $v$ = 5 a.u. and
(c) $v$ = 8 a.u. for a nanotube in vacuum and encapsulated by
SiO$_\mathrm{2}$. Since protons with impact parameters
$|x_\mathrm{0}| \geq$ 11 a.u. have enough transversal energy to make
oscillations between the opposite sides of the nanotube wall, the
results shown in Fig. \ref{fig10} resemble the motion of a particle
in a box with rigid walls owing to our use of the Doyle-Turner
potential. The amplitude of the oscillations incrises slightly due
to weakening of the image forces with increasing proton speed. More
importantly, we see very little influence of the dielectric
surrounding on the type of trajectories shown in Fig. \ref{fig10},
even at higher proton speeds or after multiple oscillations of
inside the nanotube.

Figure \ref{fig11} shows the dependence of the proton deflection
angle $\Theta_\mathrm{x}$ multiplied by the proton speed $v$ on the
dwell time $L/v$ for two proton impact parameters, $x_\mathrm{0}$ =
3 a.u. and $x_\mathrm{0}$ = 8 a.u, and three protons speed (a) $v$ =
3 a.u., (b) $v$ = 5 a.u. and (c) $v$ = 8 a.u., for a nanotube in
vacuum and encapsulated by SiO$_\mathrm{2}$. Protons with impact
parameters $|x_\mathrm{0}| \leq$ 11 a.u. do not have enough
transversal energy to bounce between the opposite sides of the
nanotube wall, but rather undergo oscillations in the transverse
plane around the minima generated by the image potential near $x
\approx$ 9.3 a.u. It is obvious from Figure \ref{fig11} that the
amplitudes of these oscillations decrease and their periods increase
with increasing proton speed, which is easily explained by the
weakening of image force as shown in Figures \ref{fig01} and
\ref{fig02}. On the other hand, such changes in the image
interaction did not give any substantial effects in Figure
\ref{fig10} because the total potential is dominated by the
repulsive interaction for proton's impact parameters $|x_\mathrm{0}|
\geq$ 11.a.u., as is obvious from Figure \ref{fig02}. However, more
important properties seen in Figure \ref{fig11} are the strong
effects of dielectric media on oscillations in the proton deflection
angle at the speeds $v$ = 5 a.u. and $v$ = 8 a.u. Namely, one
notices in Figures \ref{fig11} (b) and (c) that the shapes of
oscillations for nanotubes in SiO$_\mathrm{2}$ do not change much
between those two speeds, whereas the periods of oscillations for
nanotubes in vacuum have much shorter periods for $v$ = 5 a.u., and
much longer periods for $v$ = 8 a.u. than those in the
SiO$_\mathrm{2}$ cases. These notions can be related to the
appearance of ZDF peaks in Figures \ref{fig09} (b) and (c), as well
as to the values of forces shown in Figures \ref{fig02} (b) and (c)
for the same systems.


\section{Concluding remarks}

We have presented  the first theoretical investigation of the
effects of dynamic polarization of the carbon atom's valence
electrons on the angular distributions of protons channeled in an
(11,~9) SWNT surrounded by different dielectric media. Proton speeds
between 3 and 10 a.u., corresponding to energies of 0.223 and 2.49
MeV, have been chosen, with the nanotube's length varied between 0.1
and 1 $\mu$m.

We have confirmed here our earlier findings that in short chiral
SWNTs, it is the dynamic image interaction that gives rise to the
rainbow effect in the angular distributions of protons channeled at
the speeds below some $v \approx$ 8 a.u. \cite{borka06}. In the
presence of dielectric media, this range of the image interaction
effects is expanded to higher proton speeds because then the SWNTs
become "transparent" to the polarization effects of the surrounding
material. On the other hand, for proton speeds below $v \approx$ 3
a.u., the image interaction is almost unaffected by the surrounding
media because of the efficient screening by the nanotube.

Specifically, we have found that the effects of dielectric media are
not only of quantitative nature in affecting the positions of the
rainbow peaks, but rather can give rise to qualitative differences
compared to the case of nanotubes in vacuum, e.g., in yielding
different numbers of rainbow peaks for the same proton speed and
nanotube length. For example, we have seen that the presence of
dielectric media in our simulations both removes (see Figure
\ref{fig03}(b)), and introduces (see Figures
\ref{fig05}-\ref{fig08}(b)) rainbow singularities in the angular
distributions of protons channeled through nanotubes. Moreover, the
type of the surrounding material has also been found to affect
details of the proton angular distributions.

Going further beyond our previous study \cite{borka06}, we have
analyzed here the zero-degree focusing (ZDF) effect and found quite
substantial differences between the cases of nanotubes in vacuum and
in dielectric media when it comes to the periods of peaks in the
proton yield as a function of dwell time through the nanotube. These
effects have been further analyzed by studying typical proton
trajectories which revealed strong effects of dielectric media.
Because it may be easy to measure such features of ZDF in future
experiments on ion channeling through carbon nanotubes, we shell
devote a separate study to the ZDF in chiral SWNTs as particularly
simple prototypes of quasi-one-dimensional channeling where the
image interaction produces rich structures in the angular
distributions of channeled ions.

All our findings indicate that it is important to carefully consider
in future simulations and experiments the role played by dielectric
media in ion channeling through carbon nanotubes at the MeV
energies. For example, the best well ordered carbon nanotubes have
been grown in porous dielectric media, such as
Al$_\mathrm{2}$O$_\mathrm{3}$, so that in any analysis of ion
channeling experiments through such structures one may not
realistically ignore the influence of dielectric media. Moreover, in
the experiment by Zhu \emph{et al.} \cite{zhu05}, angular
distributions of ions channeled through the nanotubes grown inside
the pores in Al$_\mathrm{2}$O$_\mathrm{3}$ were compared with those
coming from ion channeling through the pores of pristine
Al$_\mathrm{2}$O$_\mathrm{3}$ \cite{zhu05}. It is therefore
important to perform comparative simulations of ion channeling
through channels in different materials, with and without carbon
nanotubes embedded in them, while taking proper account of the
dynamic polarization effects. While a detailed study of this problem
is in progress, some qualitative predictions can be made here for
channels inside conducting materials. Based on recent calculations
of the dynamic polarization of such structures
\cite{aris01aris01,toke00,toke01,mowb06}, it follows that the image
force would be expected to play similar role as described in the
present work, at least judging by its magnitudes for a nanotube in
vacuum and for an empty channel in metal having the same radius as
nanotube (compare thin solid lines with dashed lines in
Fig.\ref{fig01}(c)). However, it is questionable whether the atomic
structure of walls in such channels would be smooth enough to
support the rainbow effect \cite{schu04schu05,toke00,toke01}. This
leaves the ZDF as probably the most robust feature associated with
the image force, which could be probed with comparable degrees of
experimental detail for ion channeling through both pristine
channels and carbon nanotubes grown in those channels. We finally
note that such studies may further elucidate dielectric properties
of carbon nanotubes in the presence of dielectric media of relevance
to nanoelectronic, such as SiO$_\mathrm{2}$.

\begin{acknowledgments}
D.B., S.P., and N.N. acknowledge support by the Ministry of Science
and Environmental Protection of Serbia, and D.B., D.J.M. and Z.L.M.
acknowledge supports by NSERC and PREA. D. B. would also like to
thank professors Giuseppe Tenti and Frank Goodman for many useful
discussions.
\end{acknowledgments}


\end{document}